\documentclass[aps,prd,showpacs,nofootinbib,amssymb,preprint]{revtex4-1}
\pdfoutput=1
\usepackage{amssymb,graphicx,color,ulem,amsmath,slashed}
\usepackage{epsfig}

\begin{document}

\title{Testing the hypothesis of vector X17 boson by \(D\) meson, Charmonium, and \(\phi\) meson decays
}

\author{Fei-Fan Lee$^{a}$, Lam Thi Thuc Uyen$^{b}$, and Guey-Lin Lin$^{b}$}
\affiliation{$^{a}$ Department of Physics, Jimei University,
361021, Xiamen, Fujian province, P. R. China \\
$^{b}$Institute of Physics, National Yang Ming Chiao Tung University, Hsinchu 30010,
Taiwan
}
\date{\today}

\begin{abstract}
The recent ATOMKI experiments provided evidence pointing towards the existence of an X17 boson in the anomalous nuclear transitions of Beryllium-8, Helium-4, and Carbon-12. In this work, we consider X17 boson contributions to the previously measured $D$ meson decays which include $D_s^{*+} \rightarrow D_s^+ e^+ e^-$, $D_s^{*+} \rightarrow D_s^+ \gamma$, $D^{*0} \rightarrow D^0 e^+ e^-$, and $D^{*0} \rightarrow D^0 \gamma$, as well as the measured decays of $\psi(2S) \rightarrow \eta_c e^+ e^-$, $\psi(2S) \rightarrow \eta_c \gamma$, $\phi \rightarrow \eta e^+ e^-$, and $\phi \rightarrow \eta \gamma$. Using the data of the above meson decays, we perform a fitting to the coupling parameters $\varepsilon_u, \varepsilon_c$, and $\varepsilon_s$ by treating the couplings $\varepsilon_u$ and $\varepsilon_c$ as independent from each other rather than assuming the generation universality $\varepsilon_u =\varepsilon_c$. It is found that the above fitting renders $|\varepsilon_c|=7.6\times 10^{-3}$, $|\varepsilon_s|=2.4\times 10^{-3}$ and a huge magnitude for $\varepsilon_u$, which is in serious tension with $\left|\varepsilon_u\right|$ determined from ATOMKI measurements. Using our fitted range for $\varepsilon_c$ and the range for $\varepsilon_d$ from  ATOMKI measurements, we predict the range for $D^{*+} \rightarrow D^{+} e^+ e^-$ decay rate.

\end{abstract}

\maketitle
 
\newpage
\section{Introduction}
 
In the search for new physics beyond the Standard Model (SM), the study of low energy processes is an effective avenue besides those focusing on either high energy or precision frontiers. There has been a persistent anomaly in the measured angular distribution of $e^{+}e^{-}$ pairs in the 18.15 MeV decay of the excited state $^{8}$Be* by the ATOMKI collaboration over the last few years~\cite{Krasznahorkay:2015iga,Krasznahorkay:2017gwn,Krasznahorkay:2017qfd,Krasznahorkay:2019lgi}. The simplest way to account for the data is to introduce a new X17 boson with a mass in the range of approximately 16 to 17 MeV. 
The de-excitation of $^{8}{\rm Be}^*$ is assumed to emit a X17 boson which then decays to the $e^{+}e^{-}$ pair as detected by the above experiments. 
It is desirable to confirm the existence of X17 boson which undoubtedly leads to new physics beyond SM. 

The results reported in~\cite{Krasznahorkay:2015iga} gives $m_{X}=16.70\pm0.35({\rm stat})\pm 0.5({\rm syst})~\rm{MeV}$ and the ratio $\Gamma (^{8}{\rm Be}(18.15)\rightarrow~ ^{8}{\rm Be} + {X})
\times {\rm BR}({X}\rightarrow e^{+}e^{-})/\Gamma (^{8}{\rm Be}(18.15)\rightarrow~ ^{8}{\rm Be} + \gamma)\equiv R_{X}=5.8\times 10^{-6}$, with a statistical significance of $6.8\sigma$. Later analysis from more $^{8}$Be experiments gave the updated results, $m_{X}=17.01\pm0.16~\rm{MeV}$ and $R_{X}=(6\pm 1)\times 10^{-6}$~\cite{Krasznahorkay:2019lgi}. This is also the most recent data for the proposed X17 boson resulting from the $^{8}$Be anomaly, regardless of the significant anomalies observed in the case of direct proton capture~\cite{Sas:2022pgm}.

In recent years, there were further studies that suggest potential connections between X17 boson and additional anomalies observed in the decays of excited $^{4}$He~\cite{Firak:2020eil,Krasznahorkay:2019lyl,Krasznahorkay:2021joi} and $^{12}$C~\cite{Krasznahorkay:2022pxs}. These anomalies, like the $^{8}$Be case, exhibit deviations from the expected decay distributions. The masses of the proposed particle, as inferred from the anomalies observed in excited $^{4}$He and $^{12}$C decays, are reported as $16.94\pm0.12({\rm stat})\pm 0.21({\rm syst})~\rm{MeV}$~\cite{Krasznahorkay:2021joi} and $17.03\pm0.11({\rm stat})\pm 0.2({\rm syst})~\rm{MeV}$~\cite{Krasznahorkay:2022pxs}, respectively. It is remarkable that the mass values for the proposed particle as extracted from various experiments involving different nuclear systems are rather similar. This hints at a unique particle that can account for decay anomalies of excited $^8$Be, $^4$He, and $^{12}$C states\footnote{We note that there existed other experiments that perform searches for X17 boson as well. The search by VNU, Vietnam~\cite{Anh:2024req} confirmed the $^{8}{\rm Be}^*$ decay anomaly while the search by MEG II collaboration at PSI did not observe the resonance~\cite{MEGII:2024urz}.   }. 

There existed studies proposing vector couplings between X17 boson and fermions as the favorable scenario for addressing the above decay anomalies~\cite{Feng:2016jff,Feng:2016ysn,Pulice:2019xel,Feng:2020mbt,Nomura:2020kcw}. 
The scenario for axial-vector or more generally the mixture of vector and axial-vector couplings were also proposed
 ~\cite{Feng:2020mbt,Kozaczuk:2016nma,DelleRose:2017xil,DelleRose:2018eic,DelleRose:2018pgm,Seto:2020jal}. There were also proposals that the above decay anomalies can be resolved or mitigated by resorting to various unaccounted effects within SM
 ~\cite{Zhang:2017zap,Aleksejevs:2021zjw,Wong:2022jeq}.    
In this article we adopt the proposal that X17 is a vector boson. We begin by writing the couplings between X17 boson and quarks of various flavors as  
\begin{eqnarray}
{\cal L}_{X(Q,q)}=e\varepsilon_{Q} X_{\mu}\left( \overline{Q}\gamma^{\mu}Q \right) + e\varepsilon_{q} X_{\mu}\left( \overline{q}\gamma^{\mu}q \right),
\label{eq_XQq}
\end{eqnarray} 
where $q$ stands for $u$, $d$ and $s$ quarks, while $Q$ represents heavy quarks. From the data of ${ }^{8} {\rm Be}$ anomaly, $R_{X}=5.8 \times 10^{-6}$, with the assumption ${\rm BR}\left(X \rightarrow e^{+} e^{-}\right)=1$, it has been established that~\cite{Feng:2016jff,Fornal:2017msy}
$\varepsilon_{u} \simeq \pm 3.7 \times 10^{-3}$ and $\varepsilon_{d} \simeq \mp 7.4 \times 10^{-3}$, where the null result of $\pi^0\to X\gamma$ search by NA48/2~\cite{NA482:2015wmo} has been used. Further refined analysis takes into account isospin effects and their mixings~\cite{Feng:2016ysn}. Such an analysis 
considers three distinct isospin scenarios: no isospin effects, isospin mixing, and isospin mixing \& breaking. Assuming that X17 is a vector boson, a statistical $\chi^2$ test based upon data from anomalous transitions observed in excited ${ }^{8} \mathrm{Be},{ }^{4} \mathrm{He}$, and ${ }^{12} \mathrm{C}$ has been performed 
in~\cite{Denton:2023gat}. Utilizing eleven measurements that include angular and width data from the anomalous transitions in ATOMKI experiments, and the null result of $\pi^0\to X\gamma$ search~\cite{NA482:2015wmo} under the assumption ${\rm BR}\left(X \rightarrow e^{+} e^{-}\right)=1$,  the above $\chi^2$ test determines the ranges of three parameters, $m_{X}, \varepsilon_{n}, \, {\rm and }\, \varepsilon_{p}$ where the latter two are related to the quark-level couplings by $\varepsilon_p=2\varepsilon_u+\varepsilon_d$ and  $\varepsilon_n=2\varepsilon_d+\varepsilon_u$. It was found that~ \cite{Denton:2023gat} $\left|\varepsilon_{u}\right| \simeq(0.5-0.9) \times 10^{-3}$ and $\left|\varepsilon_{d}\right| \simeq(2.5-2.9) \times 10^{-3}$ with $\varepsilon_u$ and $\varepsilon_d$ in opposite signs. Specifically, the most favored parameter set is $(\varepsilon_u, \varepsilon_d)=(\pm 5\times 10^{-4}, \mp 2.9\times 10^{-3})$ and $m_{X}=16.84 \ {\rm MeV}$ if isospin effects are not considered. With both isospin mixing and breaking effects considered, the most favored parameter set becomes $(\varepsilon_u, \varepsilon_d)=(\pm 9\times 10^{-4}, \mp 2.5\times 10^{-3})$ and  $m_{X}=16.85 \ {\rm MeV}$.
It is noteworthy that the above fitted values for $\left|\varepsilon_u\right|$ and $\left|\varepsilon_d\right|$ are much smaller than those determined earlier.

Given the hypothesis of X17 boson to account for the anomalous decays of ${ }^{8} \mathrm{Be},{ }^{4} \mathrm{He}$, and ${ }^{12} \mathrm{C}$ excited states, it is of interest to test this hypothesis on
heavy hadrons which involve the couplings of X17 boson to the second and the third generation of quarks, in addition to the couplings $\varepsilon_u$ and $\varepsilon_d$. In fact, $^{8}{\rm Be}$ anomaly already motivated the study of X17 boson effect to $K^+\to \mu^+\nu_{\mu} e^+e^-$ decay through the interaction/decay chain, $K^+\to \mu^+\nu_{\mu}X\to \mu^+\nu_{\mu}e^+e^-$~\cite{Chiang:2016cyf} with the assumption $\varepsilon_s=\varepsilon_d$.
It was also proposed that X17 boson can be searched by BESIII or Belle II 
by the decay chain $J/\Psi \to \eta_c X^*\to \eta_c e^+e^-$~\cite{Ban:2020uii},
which is an excess to the regular decay chain $J/\Psi \to \eta_c \gamma^*\to \eta_c e^+e^-$.  
Furthermore, effects of X17 boson to decays of $D^{*(+,0)}$, $D^{*+}_{s}$, and the corresponding $B$ mesons were also studied in~\cite{Castro:2021gdf}.  In the latter work, it was stated that the theoretical prediction of $R_{D_s^{*+}}\equiv \Gamma(D_s^{*+} \rightarrow D^+_s e^{+} e^{-})/ \Gamma(D_s^{*+} \rightarrow D^+_s \gamma)$ 
with the contribution $D_s^{*+}\to D^{+}_s+X\to D^{+}_s +e^+ + e^-$ included in the numerator is consistent with the CLEO measurement  $R_{D_s^{*+} }=\left(7.2_{-1.6}^{+1.8}\right) \times 10^{-3}$~\cite{CLEO:2011mla}. 
We note that the authors of Ref.~\cite{Castro:2021gdf} calculated the X17 contribution to $R_{D_s^{*+}}$ by taking  $\varepsilon_{u} = \pm 3.7 \times 10^{-3}$ and $\varepsilon_{d} = \mp 7.4 \times 10^{-3}$
~\cite{Feng:2016jff} and assuming $\varepsilon_s=\varepsilon_d$, and $\varepsilon_c=\varepsilon_u$.
Furthermore, the signals due to X17 boson in other decay modes of $D^*$ and $B^*$ mesons were predicted.  

In this article, we shall revisit the contributions of X17 boson to $D^{*0}$ and $D^{*+}_{s}$ meson decays since there existed a measurement on $R_{D^{*0}}\equiv \Gamma(D^{*0} \rightarrow D^0 e^{+} e^{-})/ \Gamma(D^{*0} \rightarrow D^0 \gamma) $ after the publication of~\cite{Castro:2021gdf}. The measurement gives  $R_{D^{*0}} = (11.08 \pm 0.90) \times 10^{-3}$~\cite{BESIII:2021vyq}. Furthermore, we are motivated to tackle this issue because the updated values for $\left|\varepsilon_{u}\right|$ and  $\left|\varepsilon_{d}\right|$~\cite{Denton:2023gat} are much smaller than those determined earlier~\cite{Feng:2016jff}. 
We found that the calculation on $R_{D_s^{*+}}$ by~\cite{Castro:2021gdf} is incorrect. After correcting the calculation, we found that the prediction of $R_{D_s^{*+}}$ is consistent with CLEO data only for the
updated values of $\varepsilon_{c} \equiv \varepsilon_{u}$ and $\varepsilon_{s} \equiv \varepsilon_{d}$. 
On the other hand, the data on $R_{D^{*0}}$ requires a much larger $\varepsilon_{u}$ with the assumption $\varepsilon_{c}=\varepsilon_{u}$.
It is obvious that the updated ranges of $\varepsilon_u$ and $\varepsilon_d$ with the assumptions $\varepsilon_c=\varepsilon_u$, and $\varepsilon_s=\varepsilon_d$ cannot simultaneously account for $R_{D_s^{*+}}$ and  $R_{D^{*0}}$. To ease the tension, we note that the contribution of X17 boson to $R_{D_s^{*+}}$ behaves approximately as $\left|\varepsilon_c+2\varepsilon_s\right|$ with $\varepsilon_c$ and  $\varepsilon_s$ in opposite signs as we shall see in Fig.~\ref{fig:coup_ucs} later. Hence, one can account for both $R_{D^{*0}}$ and  $R_{D_s^{*+}}$ by raising both $\left|\varepsilon_c\right|$ and $\left|\varepsilon_s\right|$
while maintaining  $\left|\varepsilon_c+2\varepsilon_s\right|$ fixed. On the other hand, this implies that the assumption $\varepsilon_s=\varepsilon_d$ is no longer appropriate. In view of this, one may lift the restriction $\varepsilon_c=\varepsilon_u$ as well. Finally, we note that the enhancement of both $\varepsilon_c$ and $\varepsilon_s$ should impact the decays of charmonium and $\phi$ meson. Therefore, we also include $\psi(2S) \rightarrow \eta_c e^+ e^-$, $\psi(2S) \rightarrow \eta_c \gamma$, $\phi \rightarrow \eta e^+ e^-$, and $\phi \rightarrow \eta \gamma$ in our analysis. We shall perform fittings to the coupling strengths $\varepsilon_u, \varepsilon_c$, and $\varepsilon_s$ with the data of  $R_{D_s^{*+}}$,  $R_{D^{*0}}$, $R_{\psi(2S)}\equiv \Gamma(\psi(2S) \rightarrow \eta_c e^+ e^-)/\Gamma(\psi(2S) \rightarrow \eta_c \gamma)$~\cite{BESIII:2022jde},
and $R_{\phi}\equiv \Gamma(\phi \rightarrow \eta e^+ e^-)/\Gamma(\phi \rightarrow \eta \gamma)$~\cite{Achasov:2000ne,CMD-2:2000qla,KLOE-2:2014hfk}. It is important to compare the favored value of $\varepsilon_u$ in this fitting to that obtained from the decay anomalies of excited $^8$Be, $^4$He, and $^{12}$C. Here, we fix the mass of X17 boson at $16.85$ MeV which is the best-fit value of $m_{X}$ obtained in~\cite{Denton:2023gat}. We adopt this simplification since various decay rates considered here are rather insensitive to $m_{X}$ over its allowed range.

This article is organized as follows. In Sec. II, we review previous studies on the X17 contributions to heavy meson decays $H^*\to H e^+ e^-$ normalized by $H^*\to H \gamma$ 
with $H^*$ representing $D^{*0}$, $D^{*+}$, and $D^{*+}_s$ vector mesons  and $H$ their pseudoscalar counterparts.  
We note that the semi-leptonic decay $H^*\to He^+ e^-$ proceeds through both $H^*\to H\gamma^* \to He^+ e^-$ and  $H^*\to HX \to He^+ e^-$. The $H^*\to H\gamma^*$ form factor is essentially $\langle H \vert J^{\rm em}_\mu   \vert H^* \rangle$ with 
$J^{\rm em}_\mu=ee_Q \bar{Q}\gamma_{\mu}Q + ee_q \bar{q}\gamma_{\mu}q$. The evaluation of this matrix element was discussed in Refs.~\cite{Cho:1992nt,Amundson:1992yp,Cheng:1992xi,Colangelo:1993zq}. It is well known that 
$ \langle H \vert  \bar{Q}\gamma_{\mu}Q  \vert H^* \rangle$ is fixed by heavy-quark (HQ) symmetry~\cite{Isgur-Wise} while various approaches are employed to compute $\langle H \vert  \bar{q}\gamma_{\mu}q  \vert H^* \rangle$. Here, we apply the model of vector meson dominance (VMD)~\cite{BKY}
proposed in Ref.~\cite{Colangelo:1993zq} for computing the matrix element of light quark current.  This choice enables the comparison of our result with that of Ref.~\cite{Castro:2021gdf} which also employs VMD model for computing 
$\langle H \vert  \bar{q}\gamma_{\mu}q  \vert H^* \rangle$. For the $H^*\to HX$ transition, the matrix element $ \langle H \vert J^{X}_\mu   \vert H^* \rangle$ with $J^{X}_\mu=e\varepsilon_{Q} \bar{Q}\gamma_{\mu}Q +e\varepsilon_{q} \bar{q}\gamma_{\mu}q$
can be easily inferred from $\langle H \vert J^{\rm em}_\mu   \vert H^* \rangle$. We shall review the details of calculating $R_{D_s^{*+}}$ and correct the error made in  Ref.~\cite{Castro:2021gdf}. 
In the same section, we shall also discuss the implication of $R_{D^{*0}}$ on values of $\varepsilon_c$ and $\varepsilon_u$. It will be demonstrated that the conditions $\varepsilon_c=\varepsilon_u$ and $\varepsilon_d=\varepsilon_s$ should be relaxed in order that the data of 
 $R_{D_s^{*+}}$ and $R_{D^{*0}}$ can be simultaneously accounted for. In Sec. III, we present the fitting to the couplings $\varepsilon_u$, $\varepsilon_s$, and $\varepsilon_c$ with the data of  $\psi(2S) \rightarrow \eta_c e^+ e^-$, $\psi(2S) \rightarrow \eta_c \gamma$, $\phi \rightarrow \eta e^+ e^-$, $\phi \rightarrow \eta \gamma$, and that of $R_{D_s^*}$ and $R_{D^{*0}}$. We also predict the range for $R_{D^{*+}}$ using the $\varepsilon_c$ range given by the above fitting and the range for $\varepsilon_d$ given by Ref.~\cite{Denton:2023gat}. Sec. IV addresses the implications of our results and the outlook.  

\section{X17 contributions to heavy meson decays $H^*\to H + e^+ + e^-$}

We consider the $1^{-} \rightarrow 0^{-}$ heavy meson transitions  $H^{*} \rightarrow H e^{+}  e^{-}$. The Lorentz-invariant matrix elements, accounting for the contribution by the intermediate photon and X17 boson, are denoted by
$-i \mathcal{M}^{\gamma}\left(H^{*} \rightarrow H e^{+} e^{-}\right)$ and $-i \mathcal{M}^{X}\left(H^{*} \rightarrow H e^{+} e^{-}\right)$, respectively. We have 
\begin{eqnarray}
-i \mathcal{M}^{\gamma}\left(H^{*} \rightarrow H e^{+} e^{-}\right)&=&T^{\gamma}_{\mu} \frac{-ig^{\mu \nu}}{q^{2}+i\epsilon}(-ie) \bar{u}\left(p_{-}\right) \gamma_{\nu} v\left(p_{+}\right),\nonumber \\
-i \mathcal{M}^{X}\left(H^{*} \rightarrow H e^{+} e^{-}\right)&=&T^{X}_{\mu} \frac{i\left(-g^{\mu \nu}+q^{\mu} q^{\nu}/m_{X}^{2}\right)}{q^{2}-m_{X}^{2}+i m_{X} \Gamma_{X}}(-ie\varepsilon_e) \bar{u}\left(p_{-}\right) \gamma_{\nu} v\left(p_{+}\right),
\label{eq:matrix_element}
\end{eqnarray} 
where $e\varepsilon_e$ is the coupling of X17 boson to $e^{\pm}$, $\Gamma_X\equiv \Gamma(X\to e^+e^-)=\frac{e^2 \varepsilon_e^2 m_X}{12 \pi} \left( 1 + \frac{2m_e^2}{m_X^2} \right) \sqrt{1 - \frac{4m_e^2}{m_X^2}}$, $T^{\gamma, X}_{\mu}$ are the $H^{*}\left(p_{H^{*}}, \epsilon_{H^{*}}\right)\to H\left(p_{H}\right) \gamma^*(q)$ and  $H^{*}\left(p_{H^{*}}, \epsilon_{H^{*}}\right)\to H\left(p_{H}\right) X(q)$ decay amplitudes, respectively, and $p_{-}$ and $p_{+}$ are the momenta of final state electron and positron, respectively.
By Lorentz invariance, we have 
\begin{eqnarray}
T^{\gamma}_{\mu}&=&\langle H(p_H) \vert J^{\rm em}_\mu   \vert H^*(p_{H^*},\epsilon_{H^*}) \rangle=ie F_{H^{*} H \gamma}(q^2) \epsilon_{\mu \rho \alpha \beta}\epsilon_{H^{*}}^{\rho} p_{H^{*}}^{\alpha} p_{H}^{\beta},\nonumber \\
T^{X}_{\mu}&=&\langle H(p_H) \vert J^{X}_\mu   \vert H^*(p_{H^*},\epsilon_{H^*}) \rangle=ie F_{H^{*} H X}(q^2) \epsilon_{\mu \rho \alpha \beta}\epsilon_{H^{*}}^{\rho} p_{H^{*}}^{\alpha} p_{H}^{\beta},
\label{eq:T_amplitude}
\end{eqnarray} 
with $q=p_{H^*}-p_H$.
The differential decay rate for $H^{*} \rightarrow H e^{+}  e^{-}$ mediated by $\gamma$ is given by
\begin{eqnarray}
\dfrac{d\Gamma^{\gamma}(H^{\ast}\rightarrow H~e^{+}e^{-})}{dq^{2}}=\dfrac{\alpha_{\rm EM}^{2}}{72\pi}F_{H^{\ast}H\gamma}^{2}(q^{2})\dfrac{1}{q^{2}}\left( 1+\dfrac{2m_{e}^{2}}{q^{2}}\right) \sqrt{1-\dfrac{4m_{e}^{2}}{q^{2}}}\dfrac{\lambda^{3/2}(m_{H^{\ast}}^{2},m_{H}^{2},q^{2})}{m_{H^{\ast}}^{3}}, 
\label{eq:differential_gamma}
\end{eqnarray} 
where $\lambda(x, y, z)=x^{2}+y^{2}+z^{2}-2 x y-2 x z-2 y z$. Similarly, the differential decay rate mediated by X17 boson is given by
\begin{eqnarray}
\dfrac{d\Gamma^{X}(H^{\ast}\rightarrow H~e^{+}e^{-})}{dq^{2}}&=&\dfrac{\alpha^2_{\rm EM}\varepsilon_e^2}{72\pi}F_{H^{\ast}HX}^{2}(q^{2})\dfrac{q^{2}}{(q^{2}-m_{X}^{2})^{2}+m_{X}^{2}\Gamma_{X}^{2}}\left( 1+\dfrac{2m_{e}^{2}}{q^{2}}\right) \sqrt{1-\dfrac{4m_{e}^{2}}{q^{2}}} \nonumber\\ 
&\times& \dfrac{\lambda^{3/2}(m_{H^{\ast}}^{2},m_{H}^{2},q^{2})}{m_{H^{\ast}}^{3}}.
\label{eq:differential_X}
\end{eqnarray}
To compare with the data, we calculate the ratio $R_{H^*}\equiv \Gamma(H^{*} \rightarrow H e^{+} e^{-})/ \Gamma(H^{*} \rightarrow H \gamma)$ with the numerator containing contributions mediated by $\gamma$ and X17 boson, respectively. To separate these two contributions, we write $R_{H^*}=R^{\gamma}_{H^*}+R^{X}_{H^*}$. 
Following Eq.~(\ref{eq:differential_gamma}) and using $\Gamma\left(H^{*} \rightarrow H \gamma\right)=\alpha_{\rm EM}F_{H^{*} H \gamma}^{2}(0) p_{\gamma}^{3}/3$, it is easy to show that
\begin{eqnarray}
R_{H^*}^{\gamma}=\int_{q_{\rm min}^{2}}^{q_{\rm max}^{2}}\dfrac{\alpha_{\rm EM}}{3\pi}\dfrac{F_{H^{\ast}H\gamma}^{2}(q^{2})}{F_{H^{\ast}H\gamma}^{2}(0)}\dfrac{1}{q^{2}}\left( 1+\dfrac{2m_{e}^{2}}{q^{2}}\right)\sqrt{1-\dfrac{4m_{e}^{2}}{q^{2}}}\dfrac{\lambda^{3/2}(m_{H^{\ast}}^{2},m_{H}^{2},q^{2})}{(m_{H^{\ast}}^{2}-m_{H}^{2})^{3}}\mbox{d}q^{2},
\label{eq:ratio_gamma}
\end{eqnarray}
where $q^2_{\rm min}=(2m_e)^2$ and $q^2_{\rm max}=\left(m_{H^{*}}-m_{H}\right)^{2}$.
Similarly, we apply Eq.~(\ref{eq:differential_X}) to obtain 
\begin{eqnarray}
R_{H^*}^{X}&=&\int_{q_{\rm min}^{2}}^{q_{\rm max}^{2}}\dfrac{\alpha_{\rm EM}\varepsilon_e^2}{3\pi}\dfrac{F_{H^{\ast}HX}^{2}(q^{2})}{F_{H^{\ast}H\gamma}^{2}(0)}\dfrac{q^{2}}{(q^{2}-m_{X}^{2})^{2}+m_{X}^{2}\Gamma_{X}^{2}}\left( 1+\dfrac{2m_{e}^{2}}{q^{2}}\right)\sqrt{1-\dfrac{4m_{e}^{2}}{q^{2}}} \nonumber\\ 
&\times&\dfrac{\lambda^{3/2}(m_{H^{\ast}}^{2},m_{H}^{2},q^{2})}{(m_{H^{\ast}}^{2}-m_{H}^{2})^{3}}\mbox{d}q^{2}. 
\label{eq:ratio_X}
\end{eqnarray}
The above equation can be simplified by applying the narrow width approximation, 
\begin{eqnarray}
\frac{1}{\left(q^{2}-m_{X}^{2}\right)^{2}+m_{X}^{2} \Gamma_{X}^{2}}=\frac{\pi}{m_{X} \Gamma_{X}} \delta\left(q^{2}-m_{X}^{2}\right).
\label{eq:nwa}
\end{eqnarray}
For the case of $D^*$ meson decays, the form factors $F_{H^{*} H \gamma}(q^2)$ and $F_{H^{*} H X}(q^2)$ were parameterized as follows~\cite{Castro:2021gdf}:
\begin{eqnarray}
F_{H^{*} H \gamma}(q^2)&=&\sqrt{\frac{m_{H^*}}{m_H}}\left(\frac{e_Q}{m_{H^*}}+\frac{e_q}{m_q(q^2)}\right), \nonumber \\
F_{H^{*} H X}(q^2)&=&\sqrt{\frac{m_{H^*}}{m_H}}\left(\frac{\varepsilon_Q}{m_{H^*}}+\frac{\varepsilon_q}{m_q(q^2)}\right),
\label{eq:form_factors}
\end{eqnarray}
where the effective light-quark mass parameter $m_q(q^2)$ can be calculated by VMD~\cite{Colangelo:1993zq}.
\begin{figure}[t]
\begin{centering}
\includegraphics[width=0.95\textwidth]{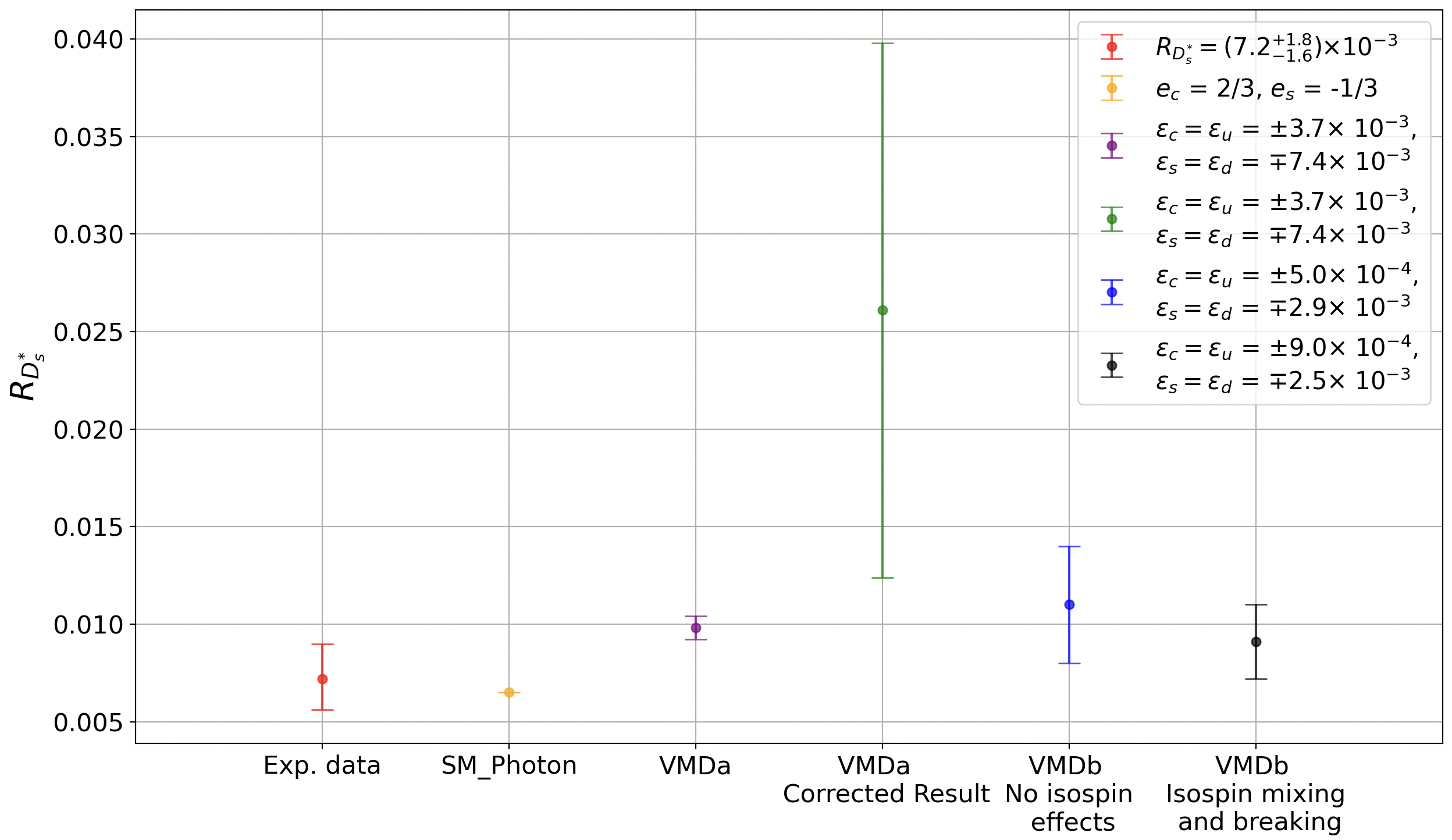}
\par\end{centering}
\caption{The experimentally measured as well as theoretically predicted values for $R_{D_s^{*+}}$. Various theoretical predictions are distinguished by different values taken for $\varepsilon_c$ and $\varepsilon_s$. VMDa stands for 
the calculation of~\cite{Castro:2021gdf} using VMD approach and employing $\varepsilon_c=\varepsilon_u$ and $\varepsilon_s=\varepsilon_d$ with $\varepsilon_{u}= \pm 3.7 \times 10^{-3}$ and $\varepsilon_{d}= \mp 7.4 \times 10^{-3}$ extracted by~\cite{Feng:2016jff,Fornal:2017msy}.  
 VMDa Corrected Result refers to our recalculation of $R_{D_s^{*+}}$ using the same values of $\varepsilon_c$ and $\varepsilon_s$. VMDb denoted in blue represents our result of using $(\varepsilon_u, \varepsilon_d)=(\pm 5\times 10^{-4}, \mp 2.9\times 10^{-3})$ which does not consider isospin effects for $^{8}{\rm Be}^*$ nuclear decays, and  that denoted in black is obtained by employing $(\varepsilon_u, \varepsilon_d)=(\pm 9\times 10^{-4}, \mp 2.5\times 10^{-3})$ where both isospin mixing and breaking effects are considered~\cite{Denton:2023gat}.
The SM prediction is denoted as SM$\_$ Photon.} 
\label{fig:R_D_s}
\end{figure}
Inferring from Eqs. (27) and (28) of Ref.~\cite{Colangelo:1993zq}, we obtain   
\begin{eqnarray}
m_{u,d}(q^2)^{-1}&=&-\sum_{V=\rho^0, \omega}\left(\sqrt{2}g_V \lambda \frac{f_V}{m_V^2}\right)\left( 1-\frac{q^2}{m_V^2}\right)^{-1} \nonumber \\
m_s(q^2)^{-1}&=&-\left(2\sqrt{2}g_V \lambda \frac{f_\phi}{m_\phi^2}\right)\left( 1-\frac{q^2}{m_\phi^2}\right)^{-1},
\label{eq:effective_mass}
\end{eqnarray}
with $\lambda$ the strength of heavy-meson coupling to light vector mesons
 described by the Lagrangian
\begin{eqnarray}
{\cal L}=i\lambda \langle {\cal H}_a\sigma_{\mu\nu}F^{\mu\nu}(\rho)_{ab} {\cal \bar{H}}_b\rangle,
\label{eq:vmd}
\end{eqnarray}
where the bracket represents taking the trace in the flavor space, ${\cal H}_a= \frac{1}{2}(1+{\slashed v})(P^*_{a,\mu}\gamma^{\mu} -P_a \gamma^5)$ is the heavy-meson field that incorporates pseudoscalar ($P_a$) and vector ($P^*_{a,\mu}$) components, 
$F_{\mu\nu}(\rho)=\partial_\mu \rho_{\nu}-\partial_\nu \rho_{\mu}+[\rho_{\mu},\rho_{\nu}]$ is the field strength tensor with $\rho_{\mu}=ig_V\hat{\rho}_{\mu}/\sqrt{2}$ where $\hat{\rho}_{\mu}$ is the $3\times 3$ matrix of the light vector meson nonet. The KSRF relations 
give $g_V=5.8$~\cite{KSRF}. The decay constant $f_V$ is defined as 
\begin{eqnarray}
\langle 0\vert \frac{1}{\sqrt{2}}(\bar{u}\gamma_{\mu}u-\bar{d}\gamma_{\mu}d)\vert \rho^0(\epsilon)\rangle &=& \epsilon_\mu f_{\rho^0},\nonumber \\
\langle 0\vert \frac{1}{\sqrt{2}}(\bar{u}\gamma_{\mu}u+\bar{d}\gamma_{\mu}d)\vert \omega(\epsilon)\rangle &=& \epsilon_\mu f_{\omega}, \nonumber \\
\langle 0\vert \bar{s}\gamma_{\mu}s \vert \phi(\epsilon)\rangle &=& \epsilon_\mu f_{\phi},
\label{eq:decay_constant}
\end{eqnarray}
where $\epsilon_{\mu}$ is the polarization of the vector meson. Using the current data of $V\to e^+e^-$~\cite{ParticleDataGroup:2024cfk}, it is found that $(f_{\rho^0}, f_\omega, f_\phi)=(0.171, 0.155, 0.232) \ {\rm GeV}^2$~\cite{Castro:2021gdf}. Finally, the parameter $\lambda$ can be determined 
by the data of semi-leptonic decays $D\to \pi(K)l\nu_l$ and $D\to K^*l\nu_l$ as pointed out in~\cite{Colangelo:1993zq}. With the updated data, it is found that~\cite{Castro:2021gdf} 
\begin{eqnarray}
\lambda=(-0.289\pm 0.016) \ {\rm GeV}^{-1}.
\label{eq:lambda}
\end{eqnarray}  

We point out that our $m_s(q^2)^{-1}$ agrees with that of~\cite{Castro:2021gdf}. On the other hand, $m_{u,d}(q^2)^{-1}$ given by Eq.~(\ref{eq:differential_X}) of~\cite{Castro:2021gdf} is twice larger than ours.
To see how this difference propagates into the final prediction, let us define the heavy and light contributions as
$H_\gamma \equiv e_Q/m_{H^*}$, $L^{(q)}_\gamma(q^2)\equiv e_q/m_q(q^2)$, $H_X\equiv \varepsilon_Q/m_{H^*}$, and $L^{(q)}_X(q^2)\equiv \varepsilon_q/m_q(q^2)$. Then, Eqs.~(\ref{eq:ratio_X}) and (\ref{eq:form_factors}) imply
\begin{eqnarray}
\frac{F_{H^*HX}(q^2)}{F_{H^*H\gamma}(0)}=\frac{H_X+L^{(u,d)}_X(q^2)}{H_\gamma+L^{(u,d)}_\gamma(0)}\,
\quad \Rightarrow \quad
R^X_{H^*}\propto \left| \frac{H_X + L_X}{H_\gamma + L_\gamma} \right|^{2}.
\label{eq:ratio_FF}
\end{eqnarray}
The factor of 2 in Ref.~\cite{Castro:2021gdf} enters through the light-quark terms in both the numerator and the denominator of the ratio $R^{X}_{H^*}$ or specifically $R^{X}_{D^*}$. This corresponds to the replacement $L^{(u,d)}_{\gamma,X}\to 2L^{(u,d)}_{\gamma,X}$ in Eq.~(\ref{eq:ratio_FF}).
 The total ratio $R_{D^*}\equiv R^{\gamma}_{D^*}+R^{X}_{D^*}$ is therefore affected.
The exact SU(3) symmetry limit amounts to setting  $f_{\rho} = f_{\omega} = f_{\phi}$ and $m_{\rho} = m_{\omega} = m_{\phi}$.  This naturally leads to another SU(3) symmetry relation $ m_u^{-1}(q^{2}) = m_d^{-1}(q^{2}) = m_s^{-1}(q^{2})$ by applying the first two relations to Eq.~(\ref{eq:effective_mass}).

Fig.~\ref{fig:R_D_s} presents the experimentally measured value of $R_{D^{*+}_s}$ as well as the corresponding theoretical predictions with and without contributions from the X17 boson.
The measured value of $R_{D^{*+}_s}$ is marked in red while the SM prediction of this ratio is marked in yellow, which considers only the photon contribution and gives $R_{D^{*+}_s}^{\gamma}=6.5 \times 10^{-3}$. This is consistent with the experimental data, $R_{D^{*+}_s}^{\text {exp}}=\left(7.2_{-1.6}^{+1.8}\right) \times 10^{-3}$~\cite{CLEO:2011mla}. 
We note that such a  consistency does not render this channel irrelevant to the X17 hypothesis. The X17 boson is motivated by anomalous nuclear transitions of excited $^8{\rm Be}$,  $^4{\rm He}$ and $^{12}{\rm C}$. Clearly, any viable X17 scenario must also be consistent with the data of electromagnetic Dalitz-type transitions in heavy-meson systems. The absence of a deviation to the SM predicted $R_{D_s^{*+}}$ in CLEO data~\cite{CLEO:2011mla} 
\footnote{The CLEO analysis~\cite{CLEO:2011mla} focuses on the integrated rate $R_{D_s^{*+}}$ and the distribution of $e^+e^-$ invariant mass $M_{ee}$. Since the latter distribution still contains large uncertainties, we only use the decay ratio $R_{D_s^{*+}}$ to test the X17 hypothesis in $D_s^{*+}\to D^{+}_s e^+ e^-$ channel.} therefore imposes a strong constraint on any new physics contribution to $ H^{*}  \to H e^{+} e^{-} $. This is precisely our main motivation for performing this cross-channel test for X17 hypothesis.

By including the contribution from X17 boson with $\varepsilon_{u} =$ $\pm 3.7 \times 10^{-3}, \varepsilon_{d} = \mp 7.4 \times 10^{-3}$~\cite{Feng:2016jff}, we obtain $R_{D^{*+}_s}=(2.6\pm 1.4)\times 10^{-2}$ as denoted by ``VMDa Corrected Result", which 
is in a mild tension with the data. For current and subsequent calculations, we take $m_X=16.85$ MeV. It is found that the decay ratios, such as  $R_{D^{*+}_s}$ and those involving other mesons, are not sensitive to $m_{X}$ over its allowed range.
Our result disagrees with the calculation by~\cite{Castro:2021gdf} which gives $R_{D_s^{*+}}=(9.82\pm 0.60)\times 10^{-3}$. The detailed comparison between two results will be given in the next paragraph.
The results denoted by VMDb are obtained with $\varepsilon_{u,d}$ taken from~\cite{Denton:2023gat}. They are divided into two cases. The first case, denoted in blue, is obtained by taking  $(\varepsilon_u, \varepsilon_d)=(\pm 5\times 10^{-4}, \mp 2.9\times 10^{-3})$ which does not consider
 isospin effects for $^{8}{\rm Be}^*$ nuclear decays.  The second case, denoted in black, is obtained by taking $(\varepsilon_u, \varepsilon_d)=(\pm 9\times 10^{-4}, \mp 2.5\times 10^{-3})$ where both isospin mixing and breaking effects are considered.
The uncertainties for theoretical predictions arise from the uncertainty of coupling parameter $\lambda$.
It is seen that both predictions under VMDb 
are consistent with the experimental data.
 
 To compare with Ref.~\cite{Castro:2021gdf} in details, 
 we note that our calculations reproduce $\varepsilon_Q/m_{H^*}=1.75\times 10^{-3} \ {\rm GeV}^{-1}$ and $\varepsilon_q/m_q(m_X^2)=-7.83\times 10^{-3} \ {\rm GeV}^{-1}$ for the decay $D_s^{*+}\to D_s^{+}X$ given in Table II of that paper.  Hence, $F_{H^{*} H X}(m_X^2)$ 
 defined by Eq.~(\ref{eq:form_factors}), is easily obtained as  $(-0.63\pm 0.05)\times 10^{-2} \ {\rm GeV}^{-1}$ rather than $(-0.91\pm 0.06)\times 10^{-2} \ {\rm GeV}^{-1}$ given in Table II of~\cite{Castro:2021gdf}.
 Here, the uncertainty of  $F_{H^{*} H X}(m_X^2)$ is due to the uncertainty of $\lambda$.  Furthermore, the quantity $e_q/m_q(0)$ in $D_s^{*+}\to D_s^+\gamma$ is related to $\varepsilon_q/m_q(m_X^2)$ by a simple scaling: $e_q/m_q(0)=e_q/\varepsilon_q\cdot \varepsilon_q/m_q(0)\simeq e_q/\varepsilon_q\cdot \varepsilon_q/m_q(m_X^2)$.  With $e_s=-1/3$, $\varepsilon_s=-7.4\times 10^{-3}$, and $\varepsilon_s/m_s(m_X^2)=-7.83\times 10^{-3} \ {\rm GeV}^{-1}$ as just mentioned, 
 one obtains $e_s/m_s(0)=-0.35 \ {\rm GeV}^{-1}$ rather than $-0.48 \ {\rm GeV}^{-1}$ given in Table I of ~\cite{Castro:2021gdf}. Since $e_c/m_{D^{*+}_s}=0.32 \ {\rm GeV}^{-1}$, one obtains $F_{D^{*+}_s D^{+}_s\gamma}(0)=-0.038\pm 0.020 \ {\rm GeV}^{-1}$ where the 
 uncertainty again arises from the uncertainty of $\lambda$. We observe that the central value of $F_{D^{*+}_s D^{+}_s\gamma}(0)$ is much smaller than that in~\cite{Castro:2021gdf} which gives  $F_{D^{*+}_s D^{+}_s\gamma}(0)=-0.17\pm 0.03 \ {\rm GeV}^{-1}$. Hence, 
 we predict a much larger $R^{X}_{D^{*+}_s}$ due to a much smaller $F^2_{D^{*+}_s D^{+}_s\gamma}(0)$ on the right hand side of  Eq.~(\ref{eq:ratio_X}). This leads to a much  larger overall ratio $R_{D^{*+}_s}$.  
%
%
\begin{table}
\caption{The values $R_{H^*}^{\gamma}$ and $R_{H^*}^{X}$ calculated with VMD model and the coupling values $\varepsilon_u=\varepsilon_c=\pm 5.0\times 10^{-4}$ and $\varepsilon_d=\varepsilon_s=\mp 2.9\times 10^{-3}$ that corresponds to best-fit parameters without considering isospin effects in $^{8}{\rm Be}^*$ decays~\cite{Denton:2023gat}.}
\begin{center}
\begin{ruledtabular}
\begin{tabular}{ccccc}
Decay mode & $R_{H^*}^{\gamma}$ & $R_{H^*}^{X}$ & $R_{H^*}$ & Experiment
\\ \hline \\
$D^{\ast 0}\rightarrow D^{0} e^+ e^- $  & $6.4\times 10^{-3}$  & $5.6\times 10^{-7}$  &  $6.4\times 10^{-3}$ & $(11.08\pm 0.90)\times 10^{-3}$~\cite{BESIII:2021vyq} \\
$D^{\ast +}\rightarrow D^{+} e^+ e^- $  & $6.4\times 10^{-3}$ & $(1.2\pm 0.4)\times 10^{-3}$  &  $(7.6\pm 0.4)\times 10^{-3}$ &  \\
$D_{s}^{\ast +}\rightarrow D_{s}^{+} e^+ e^- $ & $6.5\times 10^{-3}$  & $(4.2\pm 3.1)\times 10^{-3}$ &  $(1.1\pm 0.3)\times 10^{-2}$& $(7.2_{-1.6}^{+1.8})\times 10^{-3}$~\cite{CLEO:2011mla} \\
\end{tabular}
\end{ruledtabular}
\end{center}
\label{Tab:R_theo}
\end{table}

\begin{figure}[t]
\begin{centering}
\includegraphics[width=0.95\textwidth]{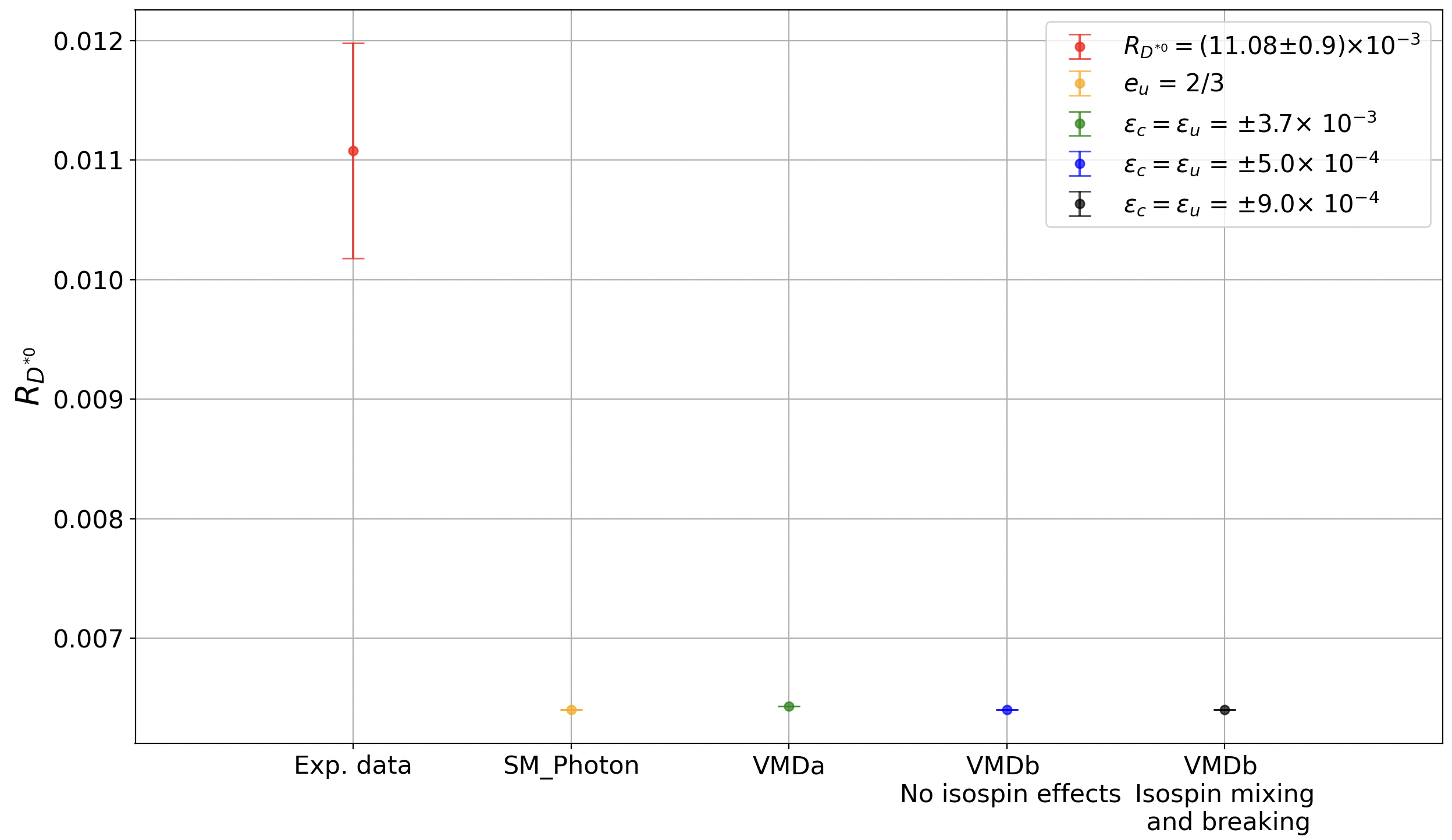}
\par\end{centering}
\caption{The experimentally measured and theoretically predicted values for $R_{D_0^*}$. 
 VMDa is the result obtained by VMD approach where $\varepsilon_c=\varepsilon_u$ and $\varepsilon_s=\varepsilon_d$ with $\varepsilon_{u}= \pm 3.7 \times 10^{-3}$ and $\varepsilon_{d}=\mp 7.4 \times 10^{-3}$~\cite{Feng:2016jff,Fornal:2017msy}. The result of VMDb denoted in blue is calculated by applying $(\varepsilon_u, \varepsilon_d) = (\pm 5\times 10^{-4}, \mp 2.9\times 10^{-3})$, and that of VMDb denoted in black is obtained by applying $(\varepsilon_u, \varepsilon_d) = (\pm 9\times 10^{-4}, \mp 2.5\times 10^{-3})$. The SM prediction is denoted as SM$\_$ Photon.}
\label{fig:R_D_0}
\end{figure}

We consider a similar analysis for the decay $D^{* 0} \rightarrow D^{0} e^{+} e^{-}$. The experimental measurement gives $R^{\text{exp}}_{D^{*0}}=(11.08 \pm 0.90) \times 10^{-3}$~\cite{BESIII:2021vyq} and it is marked in red in Fig.~\ref{fig:R_D_0}. The  SM contribution due to the photon mediation yields  $R_{D^{*0}}^{\gamma}=6.4 \times 10^{-3}$.
The full contributions $R_{D^{*0}}\equiv R_{D^{*0}}^{\gamma}+R_{D^{*0}}^{X}$  corresponding to different ranges of new physics couplings are labeled as VMDa and VMDb, respectively. Table~\ref{Tab:R_theo} presents results of $R_{H^*}$ arising from intermediate photon and X17 boson contributions for three $D^*$ meson decay modes within the VMD model, employing $\varepsilon_u=\varepsilon_c=\pm 5.0\times 10^{-4}$ and $\varepsilon_d=\varepsilon_s=\mp 2.9\times 10^{-3}$ that corresponds to best-fit parameters without isospin effects in $^{8}{\rm Be}^*$ decays considered~\cite{Denton:2023gat}. Considering isospin mixing and breaking effects in $^{8}{\rm Be}^*$ decays, i.e., $\varepsilon_u=\varepsilon_c=\pm 9.0\times 10^{-4}$ and $\varepsilon_d=\varepsilon_s=\mp 2.5\times 10^{-3}$, the corresponding $R_{H^*}$ values are $6.4\times 10^{-3}$, $(7.2\pm 0.3)\times 10^{-3}$,
and $(9.1\pm 1.9)\times 10^{-3}$ for $D^{*0}$, $D^{*+}$, and $D^{*+}_s$ decays, respectively.
 
It is seen that the SM predictions to $R^{\gamma}_{H^*}$ for $H^*=D^{*0}$, $D^{*+}$, and $D_s^{*+}$ as shown in Table~1 are almost identical. We stress that they do not arise from imposing SU(3) symmetry. In fact, we use physically measured vector meson masses and decay constants in our calculations instead of imposing SU(3) symmetry. To see this, we refer to Eq.~(\ref{eq:ratio_gamma}).
The prefactor $\alpha_{\rm EM}/(3\pi)$ and the $e^{+}e^{-}$ conversion kernel $(1/q^{2})\,(1+2m_e^{2}/q^{2})\,\sqrt{1-4m_e^{2}/q^{2}}$ are identical for all channels. 
Accordingly, any channel-to-channel differences can arise only from (i) the value of the integration upper limit $q_{\max}^{2}\equiv (m_{H^{*}}-m_H)^{2}$, 
(ii) the form-factor shape $F_{H^{*}H\gamma}^{2}(q^{2})/F_{H^{*}H\gamma}^{2}(0)$, 
and (iii) the kinematic factor $K(q^2) \equiv \lambda^{3/2}\!\left(m_{H^*}^2, m_H^2, q^2\right)/\left(m_{H^*}^2 - m_H^2\right)^3$
with $\lambda(x, y, z)=x^{2}+y^{2}+z^{2}-2 x y-2 x z-2 y z$. 
First, 
the value for $q_{\max}^{2}$ only differs by a few percent among $D^{*+}$, $D^{*0}$, and $D_s^{*+}$~\cite{ParticleDataGroup:2024cfk}. 
Second, with  $q_{\max}^{2}\sim 0.02 \ {\rm GeV}^2$ and $m_{V}^{2}\sim 0.6$--$1.0~\mathrm{GeV}^{2}$~\cite{ParticleDataGroup:2024cfk}, $q^{2}/m_{V}^{2}\le q_{\max}^{2}/m_{V}^{2} \ll 1$. This implies $F_{H^{*}H\gamma}(q^{2})/F_{H^{*}H\gamma}(0) = 1 + \mathcal{O}\!\left(\ q^2 / m_V^2\right)$, i.e., the SU(3) flavor-dependent contributions are of the order $q^2/m_V^2$, which is in the range of $(2-3)\%$. Hence, the SU(3) breaking effects here are also at the same level. 
Third, the kinematic factor $K(q^2)$ is a monotonic decreasing function of $q^2$ and its value at $q^2=q^2_{\rm max}$ vanishes. $K(q^2)$ therefore remains positive definite in the allowed kinematic range. Taking $q^2=0.01$ GeV$^2$ as an illustrative example, we find $K(q^2)=0.358, \ 0.349$, and 0.329 for $H^*=D^{*0}$, $D^{*+}$, and $D_s^{*+}$, respectively. The differences among these factors are no more than $8\%$. For other $q^2$ values, the SU(3) breaking effects remain small. 
%
In summary, we have elaborated why SU(3) breaking effects are small in $R^{\gamma}_{H^*}$ for $H^*=D^{*0}$, $D^{*+}$, and $D_s^{*+}$.  

From Fig.~\ref{fig:R_D_0},
one can see that the observed $R_{D^{*0}}$ is much larger than every theoretical prediction.  This implies that the intermediate X17 boson does not significantly enhance  $D^{* 0}\to  D^{0} e^{+} e^{-}$  decay  to match with the experimental data. The most straightforward remedy for the problem is by increasing the value of $\varepsilon_{c}$. 
However, an overly large $\varepsilon_{c}$ can easily disrupt the agreement between theory and experiment on $R_{D^*_s}$. 
To circumvent this problem, the value of $\varepsilon_{s}$ must also be adjusted accordingly. It is clear that we have to remove the assumption of $\varepsilon_{c}=\varepsilon_{u}$ and $\varepsilon_{s}=\varepsilon_{d}$.
%

We observe that the above adjustments of $\varepsilon_{c}$ and $\varepsilon_{s}$ naturally affect the decays of Charmonium ($c\bar{c}$) and $\phi$ ($s\bar{s}$) mesons. Hence, the determination of these coupling parameters should involve the data of these decay channels. 

\section{X17 couplings to charm and strange quarks}
This section aims to incorporate the decay channels $\psi(2 S) \rightarrow \eta_{c} e^{+} e^{-}$and $\phi \rightarrow \eta e^{+} e^{-}$ for determining X17 couplings to charm and strange quarks. 
We use VMD approach to calculate the normalized transition form factor (TFF) $\mathcal{F}_{H^* H V}\left(q^{2}\right)$ 
\begin{eqnarray}
\mathcal{F}_{H^* H V}\left(q^{2}\right)=\frac{F_{H^{*} H V}\left(q^{2}\right)}{F_{H^{*} H \gamma}(0)},
\label{eq:TFF}
\end{eqnarray}
with $V=\gamma, X$.
From Eq.~(\ref{eq:T_amplitude}), we have
 \begin{eqnarray}
ee_c\langle \eta_c(p_f) \vert \bar{c}\gamma_{\mu} c   \vert \psi (2S)(p_{i},\epsilon_i) \rangle&=&ie F_{\psi (2S) \eta_c \gamma}(q^2) \epsilon_{\mu \rho \alpha \beta}\epsilon_{i}^{\rho} p_{i}^{\alpha} p_f^{\beta},\nonumber \\
e\varepsilon_c\langle \eta_c(p_f) \vert \bar{c}\gamma_{\mu} c   \vert \psi (2S)(p_{i},\epsilon_i) \rangle&=&ie F_{\psi (2S) \eta_c X }(q^2) \epsilon_{\mu \rho \alpha \beta}\epsilon_{i}^{\rho} p_{i}^{\alpha} p_f^{\beta},
\label{eq:charmonium}
\end{eqnarray} 
with $q=p_i-p_f$.
It is clear that
\begin{eqnarray}
\mathcal{F}_{\psi(2S)\eta_{c} X}\left(q^{2}\right)=\left(\frac{\varepsilon_{c}}{e_c}\right)\times \mathcal{F}_{\psi (2S) \eta_{c} \gamma}\left(q^{2}\right)=3\varepsilon_c\mathcal{F}_{\psi (2S) \eta_{c} \gamma}\left(q^{2}\right)/2.
\label{eq:psi}
\end{eqnarray}
$\mathcal{F}_{\psi(2S) \eta_{c} \gamma}\left(q^{2}\right)$ is the normalized TFF for a transition between a $\psi(3686)$ and $\eta_{c}$ meson mediated by a virtual photon with the momentum $q$. It has been characterized by the pole approximation as~\cite{Gu:2019qwo}
\begin{eqnarray}
\mathcal{F}_{\psi (2S) \eta_{c} \gamma}\left(q^{2}\right)=\frac{1}{1-q^{2} / \Lambda_{\psi (2S) \eta_{c}}^{2}}.
\label{eq:pole_approx}
\end{eqnarray}
The VMD assumption has been phenomenologically successful in explaining TFF behaviors in similar decays. The effective pole mass refers to the mass of the vector meson resonance that is close to the energy scale of the decaying particle. In the analysis of $\psi(2 S) \rightarrow \eta_{c} e^{+} e^{-}$, 
$\Lambda_{\psi (2S) \eta_{c}}$ corresponds to the mass of a higher excited state, $m_{\psi(3770)}=(3773.7 \pm 0.4)\,  \mathrm{MeV/c^2}$~\cite{Gu:2019qwo}. This framework has provided a coherent explanation for the experimental observations in the Charmonium decays. 
The ratio $R_{\psi (2 S)}^{X}$
between the decay rate of $\psi (2S)\rightarrow \eta_c X\rightarrow \eta_{c} e^{+} e^{-}$ and that of $\psi (2S) \rightarrow \eta_{c}\gamma$ is given by
\begin{eqnarray}
R_{\psi (2S)}^{X}\equiv \frac{\Gamma^{X}(\psi (2S)\to \eta_c e^+e^-)}{\Gamma(\psi (2S)\to \eta_c\gamma)}\approx \frac{\Gamma(\psi (2S)\to \eta_c X)\times {\rm BR}(X\to e^+e^-)}{\Gamma(\psi (2S)\to \eta_c\gamma)},
\label{eq:Rc}
\end{eqnarray}
where the narrow width approximation for the X17 boson propagator has been used. With ${\rm BR}(X\to e^+e^-)=1$, we have $R^{X}_{\psi(2S)}=\Gamma (\psi (2S)\to \eta_c X)/\Gamma (\psi (2S)\to \eta_c \gamma)$. Applying Eq.~(\ref{eq:T_amplitude}), we arrive at  
\begin{eqnarray}
R_{\psi (2S)}^{X}=\frac{F^2_{\psi(2S)\eta_c X}(m_X^2)}{F^2_{\psi(2S)\eta_c \gamma}(0)}\times \frac{\lambda^{1/2}(m_{\psi (2S)}^2,m_{\eta_c}^2,m_X^2)}{\lambda^{1/2}(m_{\psi (2S)}^2,m_{\eta_c}^2,0)}.
\label{eq:Rc_2}
\end{eqnarray}
From Eq.~(\ref{eq:TFF}), we have $F^2_{\psi(2S)\eta_c X}(m_X^2)/F^2_{\psi(2S)\eta_c \gamma}(0)=\mathcal{F}^2_{\psi(2S)\eta_c X}(m_X^2)$. Applying Eqs.~(\ref{eq:psi}) and (\ref{eq:pole_approx}), we obtain
$\mathcal{F}^2_{\psi(2S)\eta_c X}(m_X^2)=(9\varepsilon_{c}^2/4)\times  (1-m_X^2 / \Lambda_{\psi (2S) \eta_{c}}^{2})^{-2}$, which approaches to $9\varepsilon_{c}^2/4$ as $m_X^2 / \Lambda_{\psi (2S) \eta_{c}}^{2}\to 0$.  Furthermore, 
$\lambda^{1/2}(m_{\psi (2S)}^2,m_{\eta_c}^2,m_X^2)/\lambda^{1/2}(m_{\psi (2S)}^2,m_{\eta_c}^2,0)\to 1$ due to $m_X^2\ll m_{\psi (2S)}^2  \ (m_{\eta_c}^2)$. We conclude that $R_{\psi (2S)}^{X}\to 9\varepsilon_{c}^2/4$. 

The measurement by BESIII  gives ${\rm BR}\left(\psi(2 S) \rightarrow \eta_{c} e^{+} e^{-}\right)=$ $\left(3.77 \pm 0.40_{\text {stat }} \pm 0.18_{\text {syst }}\right) \times 10^{-5}$~\cite{BESIII:2022jde}. Utilizing the branching ratio ${\rm BR}\left(\psi(2 S) \rightarrow \eta_{c} \gamma\right)=(3.6 \pm 0.5) \times10^{-3}$~\cite{ParticleDataGroup:2024cfk}, one obtains the ratio $R^{\rm exp}_{\psi(2S)}=(1.1\pm 0.3)\times 10^{-2}$ as reported in~\cite{BESIII:2022jde}.
The theoretical prediction for the photon-mediated decay rate given by the VMD model is $R_{\psi(2S)}^{\gamma}=8.9\times 10^{-3}$. The tiny difference between $R_{\psi(2S)}^{\gamma}$ and $R^{\rm exp}_{\psi(2S)}$ strongly constrain
 $R_{\psi (2S)}^{X}$, i.e., the coupling parameter $\varepsilon_c$.

The decay $\phi \rightarrow \eta e^{+} e^{-}$is a transition from a pure $\bar{s} s$ state to a mixed state involving up ($\bar{u} u$), down ($\bar{d} d$), and strange ($\bar{s} s$) quark pairs, emitting an electron-positron pair in the process. 
The measurements by SND~\cite{Achasov:2000ne}, CMD-2~\cite{CMD-2:2000qla} and KLOE-2~\cite{KLOE-2:2014hfk} yield the branching ratio for $\phi \rightarrow \eta e^{+} e^{-}$ as $\left(1.19 \pm 0.19_{\text {stat. }} \pm\right.$
$\left.0.07_{\text {syst. }}\right) \times 10^{-4},\left(1.14 \pm 0.10_{\text {stat. }} \pm 0.06_{\text {syst. }}\right) \times 10^{-4}$ and $\left(1.075 \pm 0.007_{\text {stat. }} \pm 0.038_{\text {syst. }}\right) \times 10^{-4}$, respectively. The VMD approach is utilized to estimate the lepton mass spectrum for the electromagnetic (EM) Dalitz decay $\phi \rightarrow \eta e^{+} e^{-}$. 
Following the similar approach as that of Eq.~(\ref{eq:charmonium}), we obtain the relation
 \begin{eqnarray}
\mathcal{F}_{\phi \eta X}\left(q^{2}\right)=\left(\frac{\varepsilon_{s}}{e_s}\right)\times \mathcal{F}_{\phi \eta \gamma}\left(q^{2}\right)=-3\varepsilon_s\mathcal{F}_{\phi \eta \gamma}\left(q^{2}\right).
\label{eq:phi}
\end{eqnarray}
Once more $\mathcal{F}_{\phi \eta \gamma}$ can be expressed as a simple pole~\cite{Faessler:1999de}:
\begin{eqnarray}
\mathcal{F}_{\phi \eta \gamma}\left(q^{2}\right)=\frac{1}{1-q^{2} / \Lambda_{\phi \eta}^{2}}.
\label{eq:Rs}
\end{eqnarray}
For the electromagnetic decay process $\phi \rightarrow \eta e^{+} e^{-}$, the pole mass in the VMD would naturally be a $s\bar{s}$ resonance that is close to the decaying $\phi$ meson in mass. Since $\phi(1680)$ is a higher-mass excited state of the $\phi$ meson, one thus takes $\Lambda_{\phi \eta}=m_{\phi(1680)}=(1680 \pm 20)\,  \mathrm{MeV} / \mathrm{c}^{2}$. In the scenario of $\mathrm{X} 17$ boson mediated $\phi \rightarrow \eta e^{+} e^{-}$ decay, i.e., $\phi\to \eta X$ with $X\to e^+e^-$, we obtain
\begin{eqnarray}
R_{\phi}^{X}=\frac{F^2_{\phi\eta X}(m_X^2)}{F^2_{\phi\eta \gamma}(0)}\times \frac{\lambda^{1/2}(m_{\phi}^2,m_{\eta}^2,m_X^2)}{\lambda^{1/2}(m_{\phi}^2,m_{\eta}^2,0)}\to 9\varepsilon_s^2.
\label{eq:Rst}
\end{eqnarray}
The above limit arises from $F^2_{\phi\eta X}(m_X^2)/F^2_{\phi\eta \gamma}(0)\to 9\varepsilon_{s}^2$ due to Eqs.~(\ref{eq:TFF}), (\ref{eq:phi}) and (\ref{eq:Rs}), and $\lambda^{1/2}(m_{\phi}^2,m_{\eta}^2,m_X^2)/\lambda^{1/2}(m_{\phi}^2,m_{\eta}^2,0)\to 1$ because of  $m_X^2\ll m_{\phi}^2  \ (m_{\eta}^2)$. 
Given the measurements on $R_{\phi}$  
from three separate experiments 
and the theoretical prediction, $R_{\phi}^{\gamma}=8.30 \times 10^{-3}$,  we are able to determine the allowed range for $R_{\phi}^{X}$. 
\begin{figure}[t]
\begin{centering}
\begin{tabular}{cc}
\includegraphics[width=0.49\textwidth]{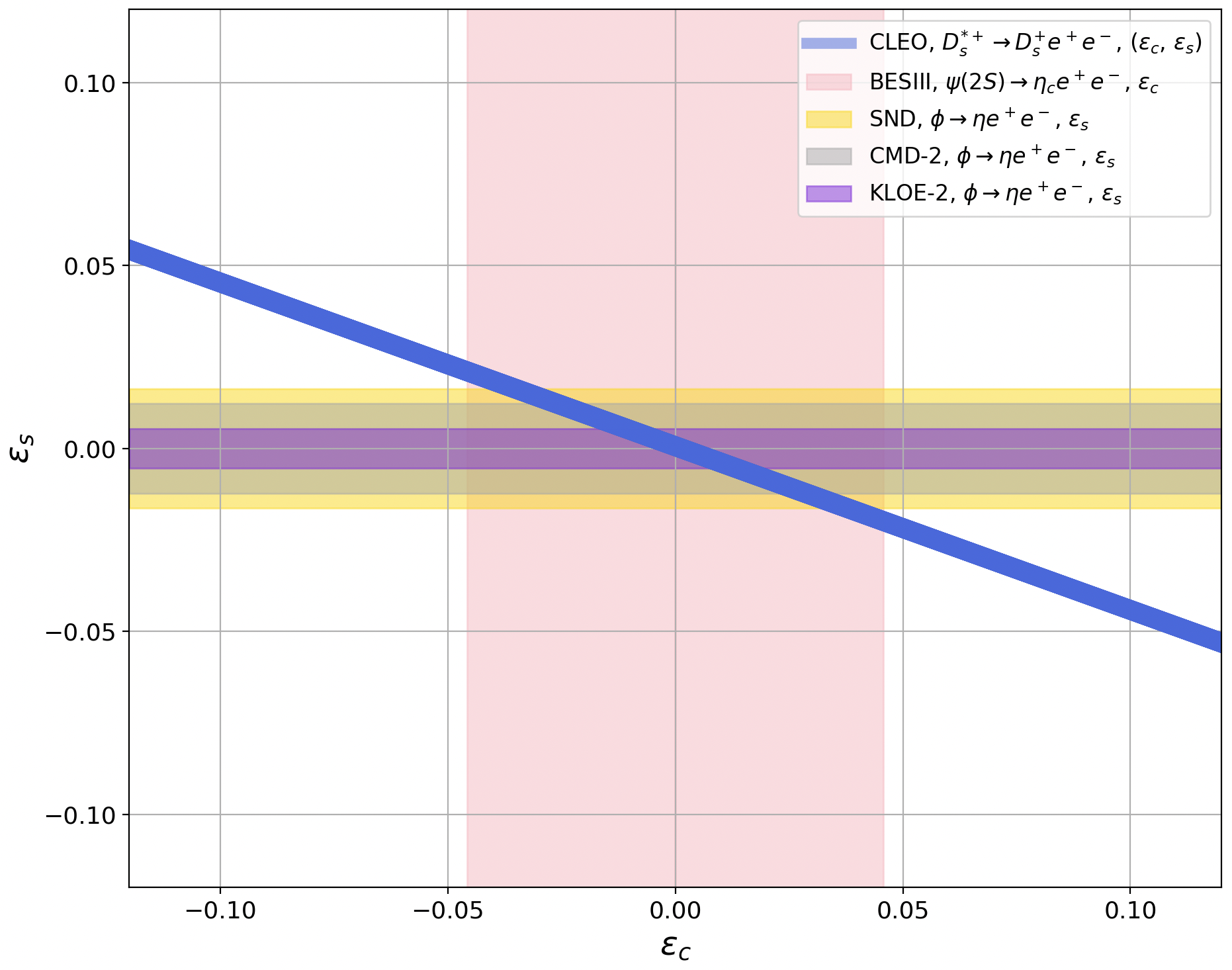} & \includegraphics[width=0.49\textwidth]{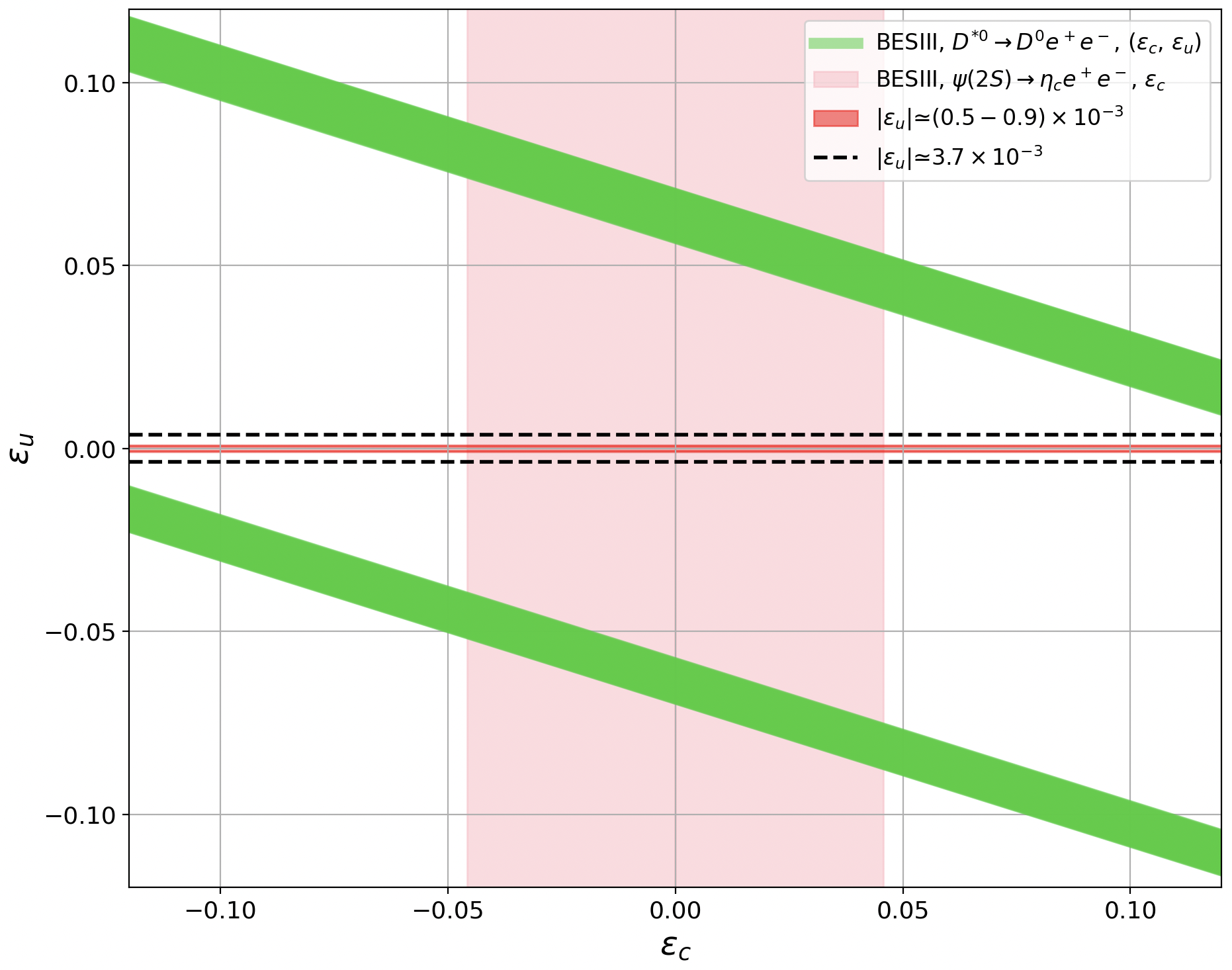}
\tabularnewline
(a)  & (b) 
\tabularnewline
\end{tabular}
\par\end{centering}
\caption{\label{fig:coup_ucs}The allowed ranges for (a) $\left(\varepsilon_{c}, \varepsilon_{s}\right)$ and (b) $\left(\varepsilon_{c}, \varepsilon_{u}\right)$ extracted by the data of $D^{*+}_s$, $D^{*0}$, $\psi(2S)$ and $\phi$ decays. The values of $\varepsilon_u$ extracted by Refs.~\cite{Feng:2016jff,Fornal:2017msy} (black dashed lines)
and Ref.~\cite{Denton:2023gat} (red band) are also shown. }
\end{figure}
%

The $1\sigma$ parameter range allowed by the ratio between $\Gamma(\psi(2S)\to \eta_c e^+ e^-)$~\cite{BESIII:2022jde} and $\Gamma(\psi(2S)\to \eta_c\gamma)$ is denoted by the pink area in Fig.~\ref{fig:coup_ucs}. The left panel depicts the allowed range for the $\left(\varepsilon_{c}, \varepsilon_{s}\right)$ pair while the right panel depicts that for $\left(\varepsilon_{c}, \varepsilon_{u}\right)$.
Furthermore,  the left panel of Fig.~\ref{fig:coup_ucs} also shows $1\sigma$ ranges for $\varepsilon_{s}$ from the data of SND~\cite{Achasov:2000ne}, CMD-2~\cite{CMD-2:2000qla} and KLOE-2~\cite{KLOE-2:2014hfk} on $\phi\to \eta e^+e^-$, 
denoted by yellow, gray, and purple areas, respectively. The blue band is the allowed range for $\left(\varepsilon_{c}, \varepsilon_{s}\right)$ by the data of $D^{*+}_s\to D^{+}_s e^+ e^-$~\cite{CLEO:2011mla}. On the right panel 
of Fig.~\ref{fig:coup_ucs}, the $1\sigma$ allowed range for $\left(\varepsilon_{c}, \varepsilon_{u}\right)$ by the data of $D^{*0}\to D^0 e^+ e^-$ is denoted by the green areas. The parameter ranges for various $D^*$ meson decays are determined with $\lambda=-0.289 \ {\rm GeV}^{-1}$. Finally, the extracted values for $\varepsilon_u $ from Refs.~\cite{Feng:2016jff,Fornal:2017msy} and Ref.~\cite{Denton:2023gat} are denoted by black dashed curves and 
red band, respectively.  
\begin{table}
\caption{The measured values and the corresponding SM predictions of four decay ratios of $D^{*0}$, $D_s^{*+}$, $\psi(2S)$, and $\phi$ mesons, respectively.   
}
\begin{center}
\begin{ruledtabular}
\begin{tabular}{ccc}
Decay Ratio  & Measured Value & SM Prediction 
\\ \hline \\
$\Gamma(D^{*0} \rightarrow D^{0} e^+ e^-)/\Gamma(D^{*0} \rightarrow D^{0}\gamma)$  &$(11.08\pm 0.90)\times 10^{-3}$~\cite{BESIII:2021vyq}  &  $6.4\times 10^{-3}$ \\
$\Gamma(D_{s}^{*+} \rightarrow D_{s}^{+} e^+ e^-)/\Gamma(D_{s}^{*+} \rightarrow D_{s}^{+}\gamma)$  & $\left(7.2_{-1.6}^{+1.8}\right) \times 10^{-3}$~\cite{CLEO:2011mla} &  $6.5\times 10^{-3}$ \\
$\Gamma(\psi(2S) \rightarrow \eta_c  e^+ e^-)/\Gamma(\psi(2S) \rightarrow \eta_c\gamma)$ & $(11 \pm 3) \times 10^{-3}$~\cite{BESIII:2022jde}  &  $8.9 \times 10^{-3}$  \\
$\Gamma(\phi\rightarrow \eta  e^+ e^-)/\Gamma(\phi\rightarrow \eta\gamma)$  & $(8.31\pm 0.33)\times 10^{-3}$~\cite{KLOE-2:2014hfk} & $8.30 \times 10^{-3}$ \\
\end{tabular}
\end{ruledtabular}
\end{center}
\label{Tab:Br_Obs}
\end{table}

\begin{figure}[t]
\begin{centering}
\includegraphics[width=0.95\textwidth]{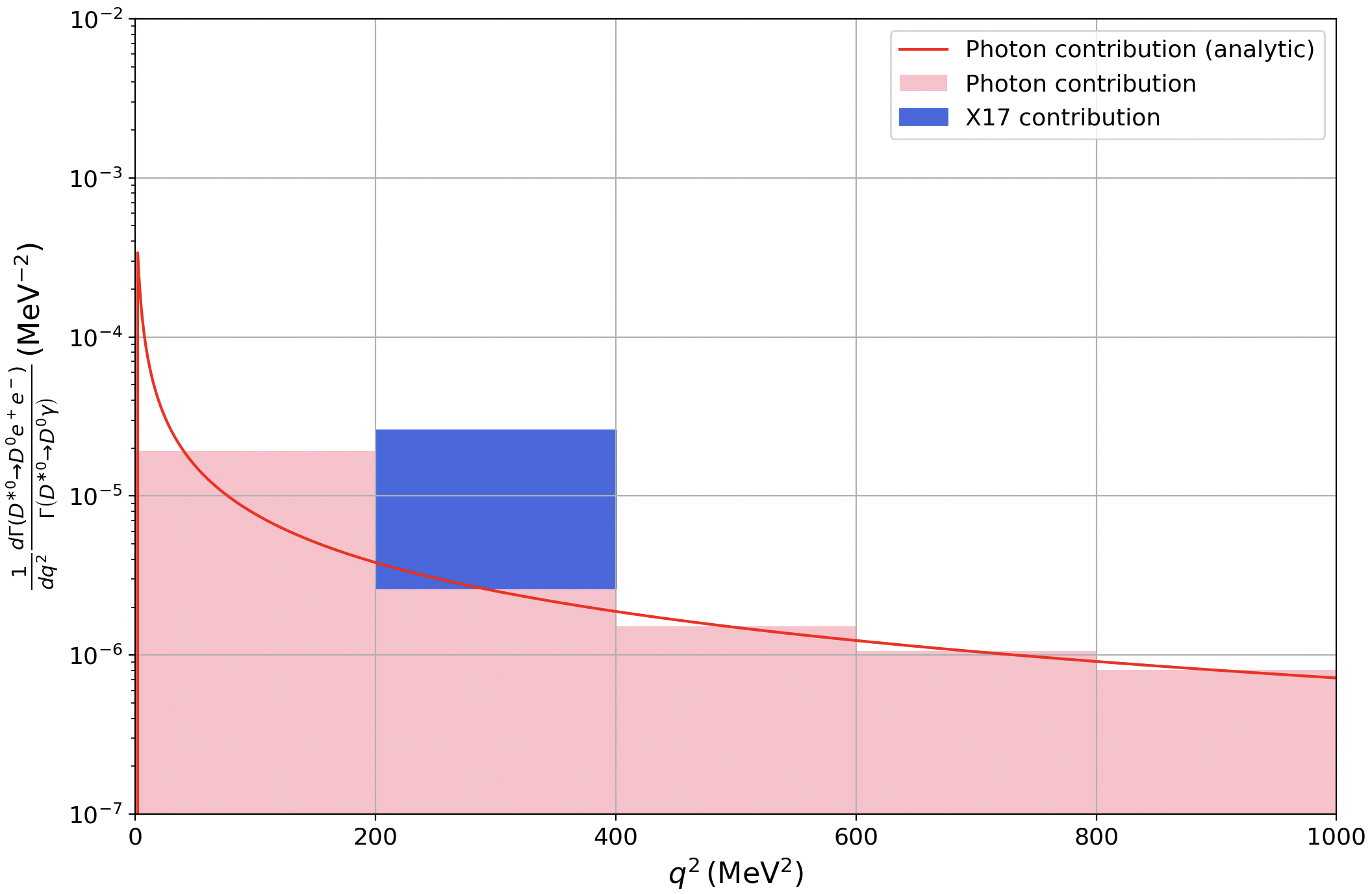}
\par\end{centering}
\caption{The distribution ${\rm d}R_{D^{*0}}/{\rm d} q^2$ as a function of $e^+e^-$ invariant mass $q^2$ with the bin width $\Delta q^2=200  \ {\rm MeV}^2$. 
The SM contribution, ${\rm d} R^{\gamma}_{D^{*0}}/{\rm d} q^2$, and X17 contribution, ${\rm d} R^{X}_{D^{*0}}/{\rm d} q^2$, with $200 \ {\rm MeV}^2$ invariant-mass binning are denoted in pink and blue, respectively.  
For comparisons, the analytic result for SM contribution ${\rm d}R^{\gamma}_{D^{*0}}/{\rm d}q^2$ is represented by the red curve.}
\label{fig:spectrum}
\end{figure}

As summarized in Table~\ref{Tab:Br_Obs}, the measured value of $R_{D^{*0}}$ is much greater than the SM predicted value. 
The difference between the two is $\delta R_{D^{*0}}=(4.68\pm 0.90)\times 10^{-3}$, which is roughly $3/4$ of $R^{\gamma}_{D^{*0}}$. If such a difference is due to X17 contribution, there should exist a peak in the distribution ${\rm d}R_{D^{*0}}/{\rm d} q^2$ at $q^2=m_X^2$.  In Fig.~\ref{fig:spectrum} we plot 
  ${\rm d}R_{D^{*0}}/{\rm d} q^2$ with the energy binning adopted by Ref.~\cite{BESIII:2021vyq}, i.e., $\Delta q^2=200 \ {\rm MeV}^2$.  The SM contribution, ${\rm d} R^{\gamma}_{D^{*0}}/{\rm d} q^2$, and X17 contribution, ${\rm d} R^{X}_{D^{*0}}/{\rm d} q^2$, with the $200 \ {\rm MeV}^2$ invariant-mass binning are denoted in pink and blue, respectively.  
 For comparisons, the analytic result for SM contribution ${\rm d}R^{\gamma}_{D^{*0}}/{\rm d}q^2$ is represented by the red curve. Since $m_X^2\approx 280 \ {\rm MeV}^2$, X17 contribution to ${\rm d} R_{D^{*0}}/{\rm d} q^2$ is at the second bin. As a result, 
  the second bin would surpass the first bin in events. This is in contrary to 
  the case with ${\rm d} R^{\gamma}_{D^{*0}}/{\rm d} q^2$ where event number in the first bin dominates that of any other bin. Although the spectral plot in~\cite{BESIII:2021vyq} does not show this feature, it remains inconclusive since it is also stated in that paper that the 
  $q^2$ distribution measurement is not yet meaningful 
 due to the limited statistics. 

 \begin{figure}[t]
\begin{centering}
\includegraphics[width=0.7\textwidth]{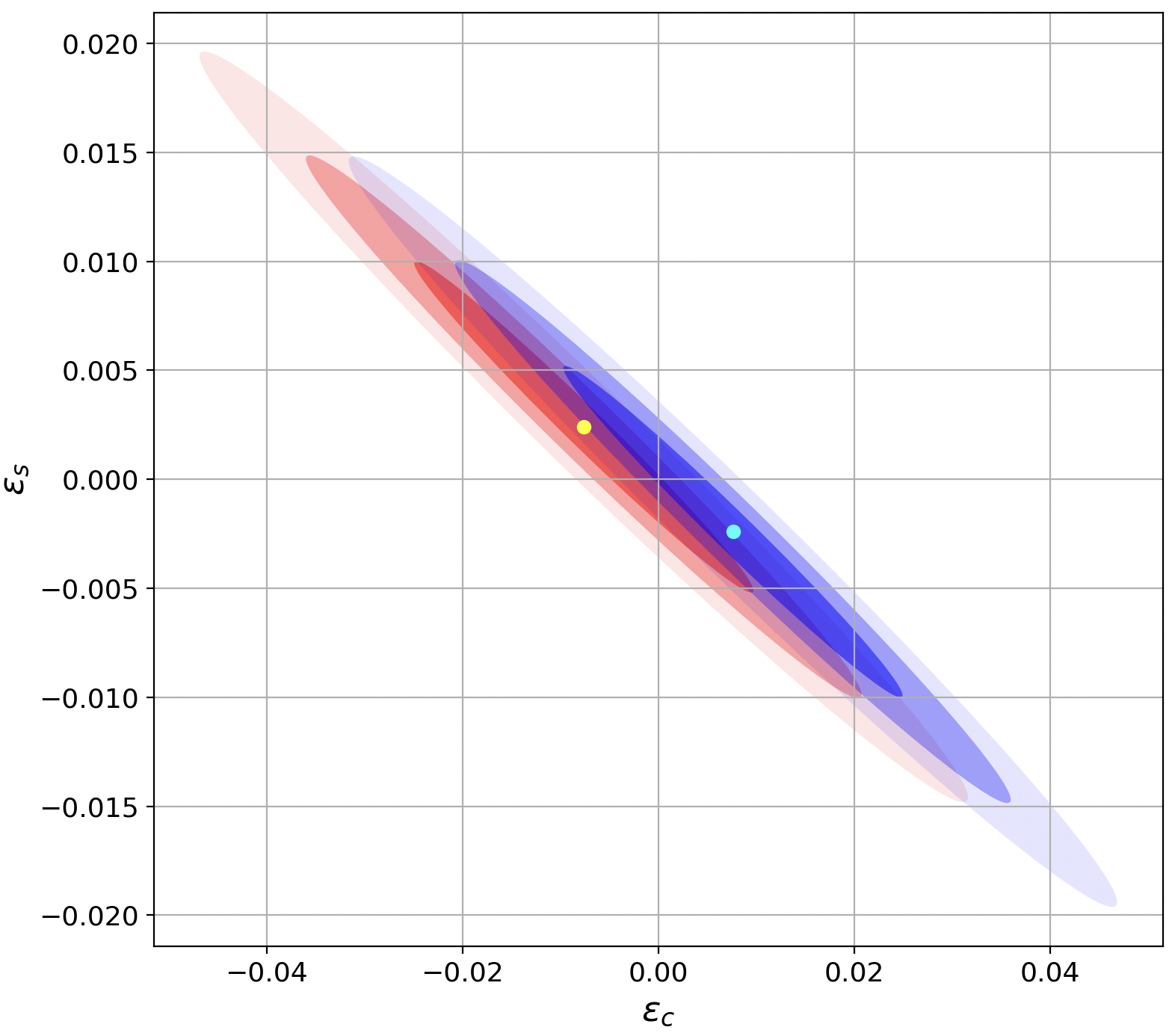}
\par\end{centering}
\caption{The best-fit values and 1, 2, and 3$\sigma$ ranges of $\varepsilon_c$ and $\varepsilon_s$  considering only the data of $R_{D_s^{*+}}$, $R_{\psi(2S)}$, and $R_{\phi}$. The best-fit values of $\varepsilon_c$ and $\varepsilon_s$ are $\pm 7.6\times 10^{-3}$ and
$\mp 2.4\times 10^{-3}$, respectively. }
\label{fig:cs_flavor}
\end{figure}

Since more statistics for $D^{*0} \rightarrow D^{0} e^{+} e^{-}$ is required for deciphering the nature of large excess in $R_{D^{*0}}$, it is of interest to perform two fittings with one that includes only $D_s^{* +} \rightarrow D_s^{+} e^{+} e^{-}$, $\psi(2S)\to \eta_c e^+e^-$, and
 $\phi\to \eta e^+e^-$ data and the other one that includes $D^{*0} \rightarrow D^{0} e^{+} e^{-}$ data as well. The first fitting only involves $\varepsilon_c$ and $\varepsilon_s$ which are not constrained by ATOMKI measurements. The result of the first fitting is shown in Fig.~\ref{fig:cs_flavor}. We employ the $\chi^{2}$ statistical test for fitting the coupling parameters $\varepsilon_{c}$, and $\varepsilon_{s}$  with the three relevant measurements given in Table~\ref{Tab:Br_Obs}.
For $\phi\to \eta e^+e^-$ decay, we only employ the data by KLOE-2~\cite{KLOE-2:2014hfk} since its uncertainty is much smaller than those of earlier measurements. We minimize the $\chi^2$ defined as  
\begin{eqnarray}
\chi^{2}=\sum_{i=1}^{3} \frac{\left(R_{i}^{\rm th}\left(\varepsilon_{c}, \varepsilon_{s}, \lambda\right)-R_{i}^{\rm ob}\right)^{2}}{\sigma_{i}^{2}}  + \frac{\left( \lambda - \lambda_{\rm best \, fit} \right)^{2}}{\sigma_{\lambda}^{2}},
\label{eq:chi_square}
\end{eqnarray}
where $R_{i}^{\rm th}$ denotes the theoretically predicted value for the $i$-th decay ratio with $i=D^{*+}_s, \ \psi(2S), \ {\rm and}  \ \phi$. These ratios depend on the parameters $\varepsilon_{c}$, $\varepsilon_{s}$, and $\lambda$. $R_{i}^{\rm ob}$ refers to the observed value for the $i$-th decay ratio with $\sigma_{i}$ the corresponding uncertainty for the measurement. 
The parameter $\lambda$ is defined in Eq.~(\ref{eq:vmd}) with its range given by Eq.~(\ref{eq:lambda}).
We hence take $\lambda_{\rm best \, fit}=-0.289$ GeV$^{-1}$ and $\sigma_\lambda=0.016$ GeV$^{-1}$.  
Our fitting uses the measured decay ratios of $D_s^{*+}$, $\psi(2S)$, and $\phi$ mesons as outlined in Table~\ref{Tab:Br_Obs}.   
We have  $\chi^2_{\rm min}=0.45$ for the fitting. 
\begin{figure}[t]
    \begin{center}
    \includegraphics[width=0.46\columnwidth]{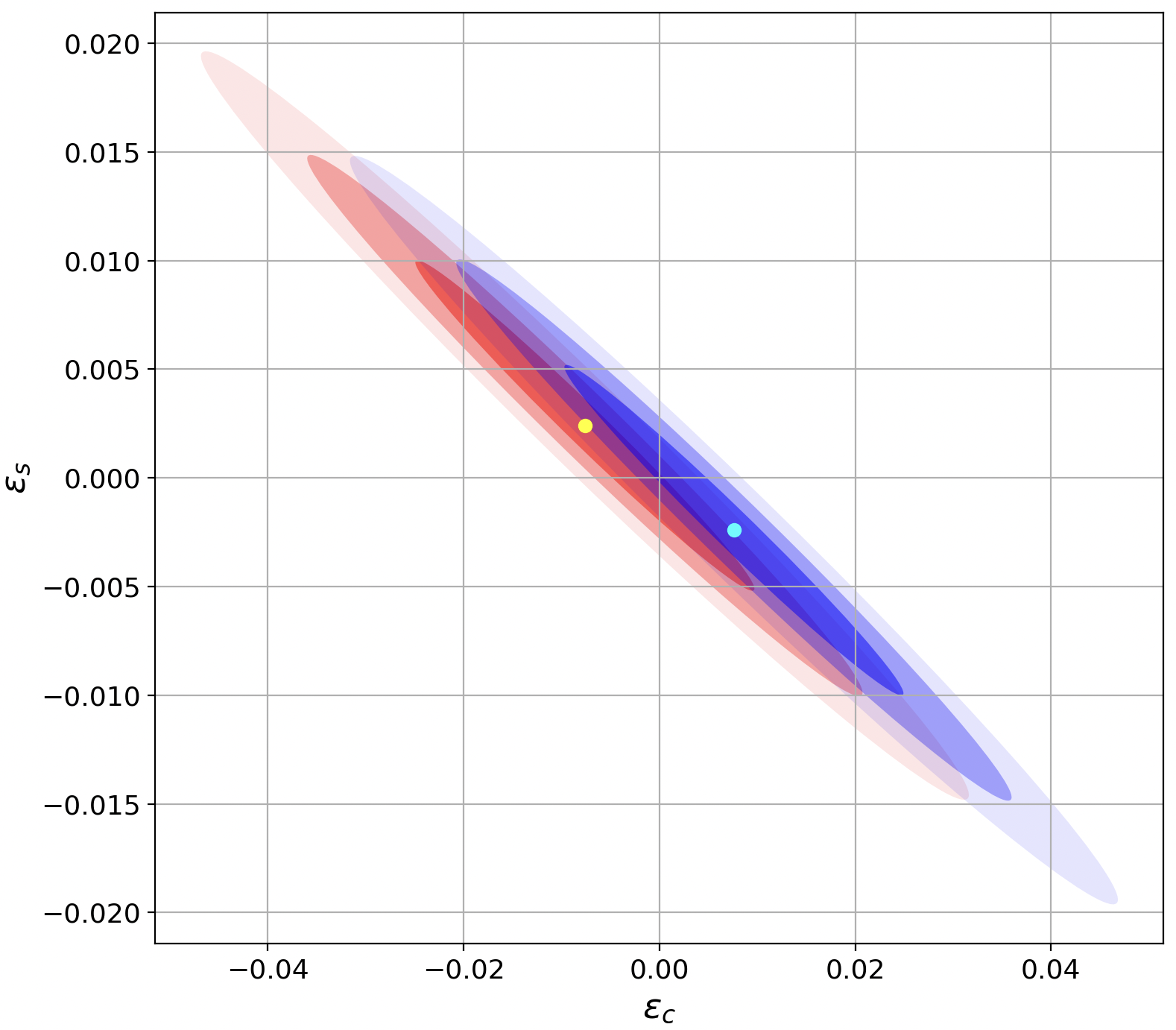}
    \includegraphics[width=0.46\columnwidth]{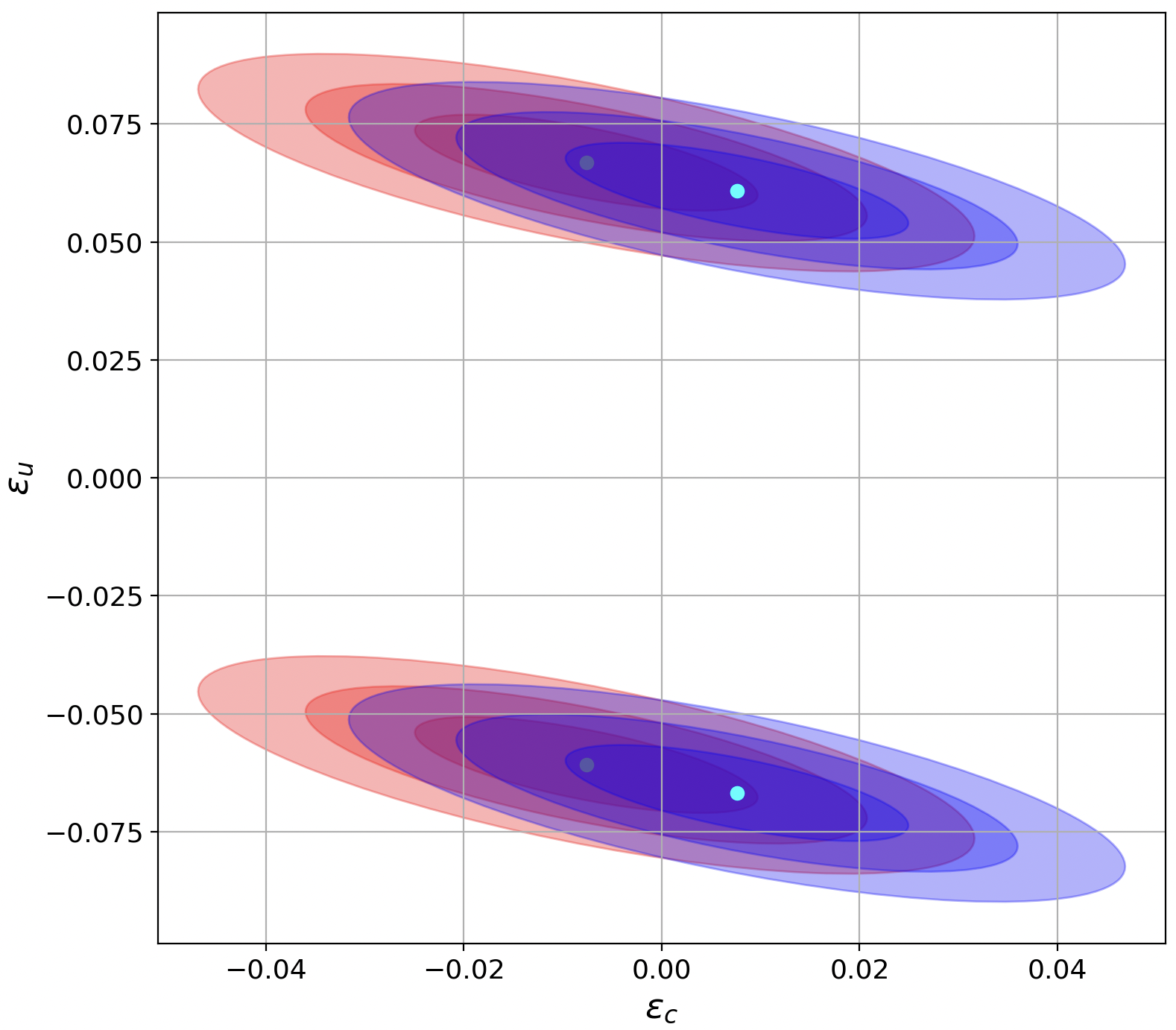}
    \includegraphics[width=0.46\columnwidth]{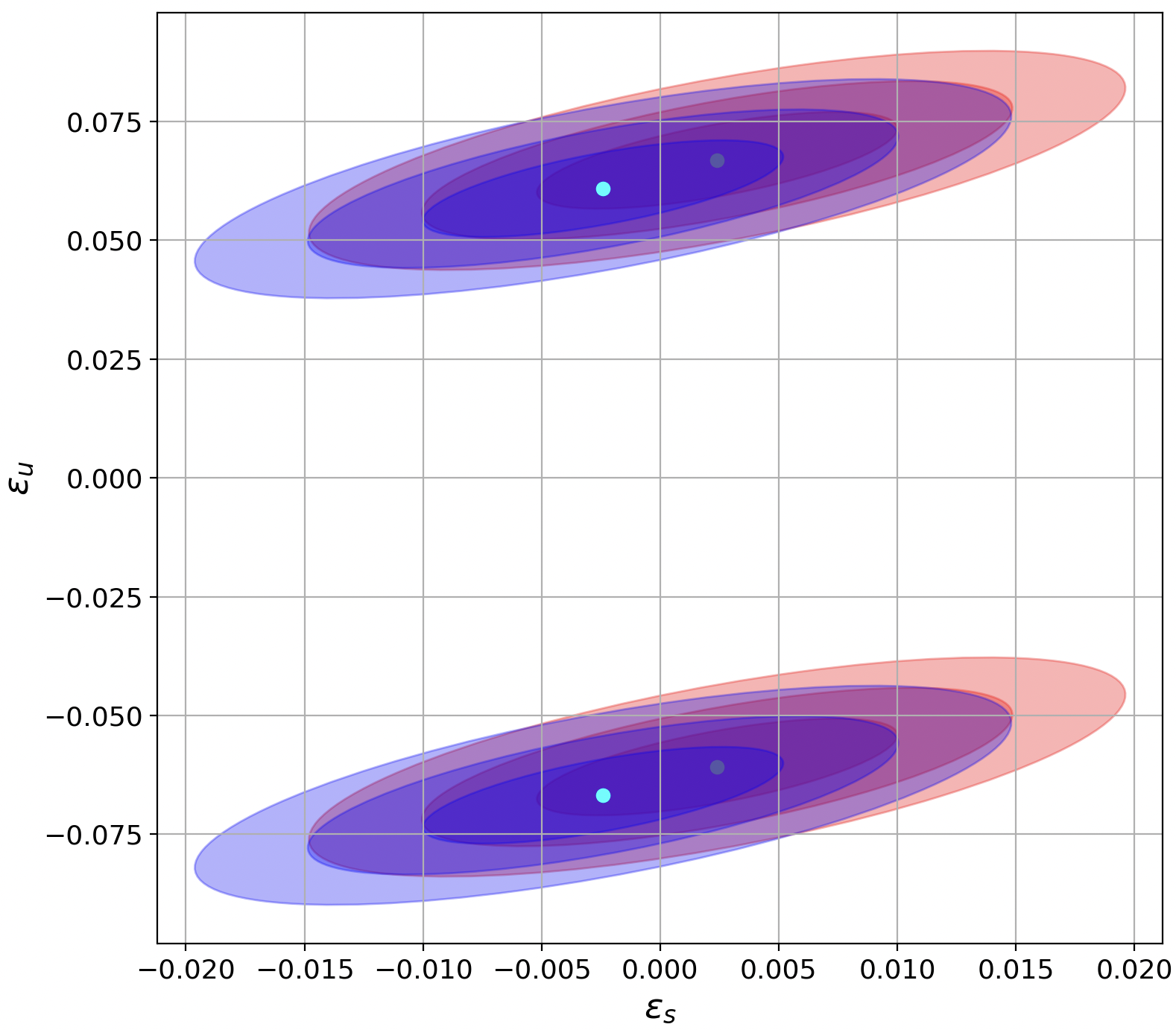}
    \caption{The best-fit values and 1, 2, and 3$\sigma$ ranges of coupling parameters projected onto  $(\varepsilon_c,\varepsilon_s)$, $(\varepsilon_c,\varepsilon_u)$, and $(\varepsilon_s,\varepsilon_u)$ planes. }
\label{fitting_range}
\end{center}
\end{figure}
The best-fit values are given by $\varepsilon_c=\pm 7.6\times 10^{-3}$ and $\varepsilon_s=\mp 2.4\times 10^{-3}$. 
We note that the best-fit value for $\varepsilon_c$ is significantly larger in magnitude than that of
 $\varepsilon_u$ obtained from fitting the ATOMKI data~\cite{Denton:2023gat} (see also the caption of Table~\ref{Tab:R_theo}). On the other hand, the allowed range for $\varepsilon_c$ is still too large to indicate whether these new physics couplings are universal in quark generations or not.

Let us now include  $D^{*0} \rightarrow D^{0} e^{+} e^{-}$ into the fitting. We note that only this decay among our considered modes involve X17 coupling to the first generation of quarks, i.e., $\varepsilon_u$.
The fitting results projected onto $(\varepsilon_c,\varepsilon_s)$, $(\varepsilon_c,\varepsilon_u)$, and $(\varepsilon_s,\varepsilon_u)$ planes are shown in Fig.~\ref{fitting_range}. 
The best-fit values for the coupling parameters are $\left|\varepsilon_{c}\right|=7.6\times 10^{-3}$ and $\left|\varepsilon_{s}\right|=2.4\times 10^{-3}$
while there are two best-fit solutions for  $\left|\varepsilon_{u}\right|$, which are $6.1\times 10^{-2}$ and $6.7\times 10^{-2}$, respectively. 
To understand the origin of double solutions for $\varepsilon_u$, let us  apply the narrow width approximation, Eq.~(\ref{eq:nwa}), to Eq.~(\ref{eq:ratio_X}) and write the theoretical prediction to $D^{*0}(c\bar u)$ decay ratio as
\begin{eqnarray}
R^{\rm th}_{D^{*0}} = R^\gamma_{D^{*0}} + \kappa\,(A\,\varepsilon_c+B\,\varepsilon_u)^2,
\label{eq: Ree_D*0}
\end{eqnarray}
here $A\equiv 1/m_{D^{*0}}$ and $B\equiv 1/m_u(m_X^2)$ with $m_u(q^2)^{-1}$ given by Eq.~(\ref{eq:effective_mass}). The coefficient 
$\kappa\equiv K/D^2$ collects the kinematic factor $K\equiv \lambda^{3/2}(m_{D^{*0}}^{2},m_{D^{0}}^{2},m_X^{2})/(m_{D^{*0}}^{2}-m_{D^{0}}^{2})^{3}$ and the photon-contribution normalization $D\equiv e_c/m_{D^{*0}}+e_u/m_u(0)$ from Eqs.~(\ref{eq:ratio_X}) and (\ref{eq:form_factors}).
For a given value for $\chi^2_{D^{*0}}(\varepsilon_u,\varepsilon_c) \equiv  \big( R_{D^{*0}}^{\text{th}}(\varepsilon_u,\varepsilon_c) - R_{D^{*0}}^{\exp} \big)^2 / \sigma_{D^{*0}}^{2}$, we apply Eq.~(\ref{eq: Ree_D*0}) and obtain $A\varepsilon_c+B\varepsilon_u=\pm\sqrt{C}$ where $C>0$ is a constant determined by
the given $\chi^2_{D^{*0}}$ value and the uncertainty $\sigma_{D^{*0}}^2$ in the measurement. We then obtain the solutions for $\varepsilon_u$ as $\varepsilon_u = (-A\varepsilon_c \pm \sqrt{C})/B$. That ``$\pm$'' leads to the existence of double solutions for $\varepsilon_u$, i.e., the fitting returns two values for $|\varepsilon_u|$ (here $|\varepsilon_u| \simeq 0.061$ and $|\varepsilon_u| \simeq 0.067$) while keeping $|\varepsilon_c| \simeq 7.6 \times 10^{-3}$ and $|\varepsilon_s| \simeq 2.4 \times 10^{-3}$. 


\begin{table}
\caption{Comparison of best-fit parameters with and without including the $R_{D^{*0}}$ data} 
\begin{center}
\begin{ruledtabular}
\begin{tabular}{ccccc}
Scenario & $\varepsilon_c$ & $\varepsilon_s$ & $\varepsilon_u$ & Comment 
\\ \hline \\
Fit without $R_{D^{*0}}$ & $\pm 7.6\times10^{-3}$ & $\mp 2.4\times10^{-3}$ & -- & -- \\
Fit with $R_{D^{*0}}$ & $\pm 7.6\times10^{-3}$ & $\mp 2.4\times10^{-3}$ &
$\pm 6.1\times10^{-2}$/$\mp 6.7\times10^{-2}$ & tension with ATOMKI \\
\end{tabular}
\end{ruledtabular}
\end{center}
\label{tab:eps_summary}
\end{table}

The fitted $\varepsilon_c$ and $\varepsilon_s$  are the same as those in the previous fitting with $R_{D^{*0}}$ excluded.
Such a result can be understood from Fig.~\ref{fig:coup_ucs}. On the left panel of the figure, the narrow bands resulting from CLEO~\cite{CLEO:2011mla} and KLOE-2~\cite{KLOE-2:2014hfk} experiments essentially determine the favored ranges for $\varepsilon_c$ and $\varepsilon_s$. When $\varepsilon_c$ is frozen to a small range as the above, the favored ranges for $\varepsilon_u$ can be seen from the right panel of Fig.~\ref{fig:coup_ucs} as the intersections of the tiny $\varepsilon_c$ favored range with the green bands dictated by the measurement on $R_{D^{*0}}$~\cite{BESIII:2021vyq}. Clearly, $R_{D^{*0}}$ data does not affect the favored range for $(\varepsilon_c,\varepsilon_s)$ unless the above intersections do not occur.
Up to the $3\sigma$ range, the value of $\left|\varepsilon_u\right|$ is of the order $10^{-2}$, which is notably larger than that determined from the anomalous decays of ${ }^{8} \mathrm{Be},{ }^{4} \mathrm{He}$, and ${ }^{12} \mathrm{C}$, i.e., $\left|\varepsilon_{u}\right| \simeq(0.5-0.9) \times 10^{-3}$~\cite{Denton:2023gat}. 
In other words, the value of $\left|\varepsilon_u\right|$ extracted from the decays of $D^{*0}$, $D^{*+}_s$, $\psi(2S)$, and $\phi$ mesons is in a serious tension with that extracted by ATOMKI measurements. The significant increase of $\left|\varepsilon_u\right|$ in this case can be attributed to the anomalously large branching ratio of  $D^{* 0} \rightarrow D^{0} e^{+} e^{-}$ measured by BESIII~\cite{BESIII:2021vyq} as illustrated by Fig.~\ref{fig:R_D_0} and the right panel 
of Fig.~\ref{fig:coup_ucs}. 
We point out that both $\varepsilon_u$ branches in Eq.~(\ref{eq: Ree_D*0}) give rise to the same  $R_{D^{*0}}^{\text{th}}$. Hence, both produce the same $\chi^2_{D^{*0}}(\varepsilon_u,\varepsilon_c)$. 
Since none of $R_{D^{*+}_s}$, $R_{\psi(2S)}$, and $R_{\phi}$ depends on $\varepsilon_u$, therefore both $\varepsilon_u$ branches also lead to the same overall $\chi^2$.
In other words, both solutions for $\varepsilon_u$ are physically acceptable. As a summary, we list the best-fit parameters obtained with and without the $R_{D^{*0}}$ data in Table~\ref{tab:eps_summary}.

\begin{figure}[t]
\begin{centering}
\includegraphics[width=0.85\textwidth]{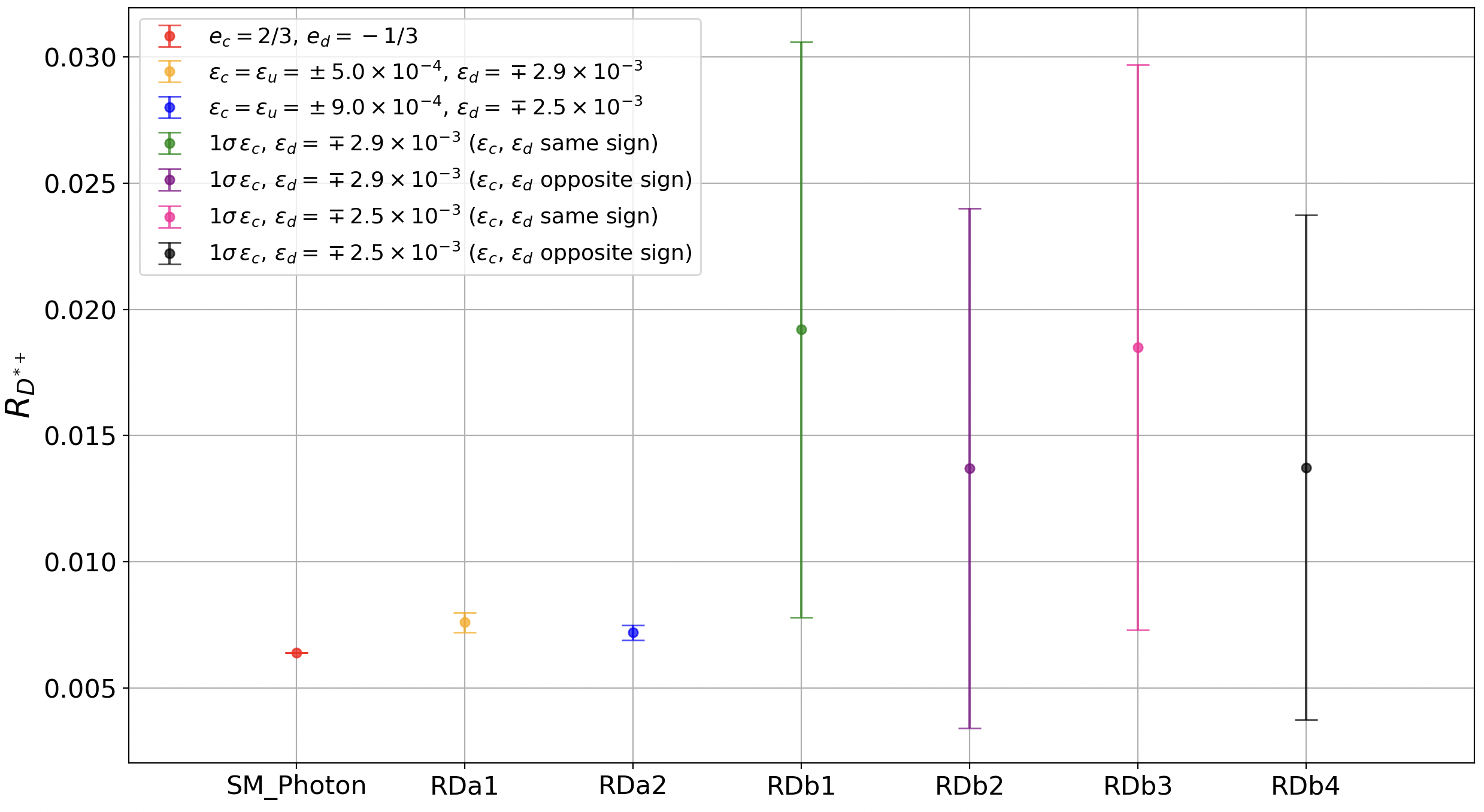}
\par\end{centering}
\caption{The theoretically predicted values for $R_{D^{*+}}$ derived from Eqs.~(\ref{eq:ratio_gamma}) and ~(\ref{eq:ratio_X}) are shown. The scenarios RDa1 and RDa2 assume $\varepsilon_c = \varepsilon_u$, where RDa1 is calculated using $(\varepsilon_u, \varepsilon_d) = (\pm 5 \times 10^{-4}, \mp 2.9 \times 10^{-3})$, and RDa2 is obtained by applying $(\varepsilon_u, \varepsilon_d) = (\pm 9 \times 10^{-4}, \mp 2.5 \times 10^{-3})$~\cite{Denton:2023gat}. The labels RDb1, RDb2, RDb3, and RDb4 represent the 1-sigma range of $\varepsilon_c$ from Fig.~\ref{fig:cs_flavor}, with different assumptions about $\varepsilon_d$. For $\varepsilon_d = \mp 2.9 \times 10^{-3}$, RDb1 corresponds to the same-sign scenario of $(\varepsilon_c, \varepsilon_d)$, while RDb2 represents the opposite-sign scenario. Similarly, for $\varepsilon_d = \mp 2.5 \times 10^{-3}$, RDb3 and RDb4 correspond to the same-sign and opposite-sign scenarios, respectively. The SM prediction is denoted as SM\_Photon.}
\label{fig:RDplus}
\end{figure}
 
With the above information on X17 coupling parameters, we give our predictions for the not-yet measured ratio $R_{D^{*+}} \equiv \Gamma(D^{*+} \to D^+ e^+ e^-)/\Gamma(D^{*+} \to D^+ \gamma)$ in Fig.~\ref{fig:RDplus}. The SM\_Photon is the Standard Model baseline with only the virtual-photon contribution in $\Gamma(D^{*+} \to D^+ e^+ e^-)$. RDa1 and RDa2 give predictions assuming the generation-universality 
$\varepsilon_c = \varepsilon_u$.  The former takes  $(\varepsilon_u, \varepsilon_d)=(\pm 5 \times 10^{-4}, \mp 2.9 \times 10^{-3})$ while the latter adopts $(\varepsilon_u, \varepsilon_d)=(\pm 9 \times 10^{-4}, \mp 2.5 \times 10^{-3})$~\cite{Denton:2023gat}.
 Both scenarios predict $R_{D^{*+}}$ above the SM value due to the additional X17 mediated amplitude. 
The vertical bands RDb1–RDb4 relax generation universality by scanning the $1\sigma$ range of $\varepsilon_c$ obtained from the combined fitting  to $R_{D_s^{*+}}$, $R_{\phi(2S)}$, and $R_{\phi}$ given in Table~\ref{Tab:Br_Obs} with $|\varepsilon_d|$ fixed at either $2.9 \times 10^{-3}$ or $2.5 \times 10^{-3}$ and different relative signs between $\varepsilon_c$ and $\varepsilon_d$ considered. 
The central value of each $R_{D^{*+}}$ prediction primarily 
depends on
the magnitude of $\varepsilon_c$, i.e., when $\vert \varepsilon_c \vert$ is larger, the X17 contribution to $\Gamma (D^{*+} \to D^+ e^+ e^-)$ 
is larger such that $R_{D^{*+}}$ is shifted upward; the opposite holds for a smaller $\vert \varepsilon_c \vert$. Meanwhile, the range of each RDb band is controlled by how $\varepsilon_c$ and $\varepsilon_d$ interferes each other. If $\varepsilon_c$ and $\varepsilon_d$ have the same sign, their associated amplitudes add up; 
if they have opposite signs, their corresponding amplitudes partially cancel each other. 
If a future measurement finds $R_{D^{*+}}$ significantly above the SM\_Photon baseline and within the RDb bands as shown in Fig.~\ref{fig:RDplus}, then the ATOMKI-sized $|\varepsilon_c| = |\varepsilon_u| \simeq (0.5–0.9) \times 10^{-3}$ would generally be too small to account for the observed enhancement. This then points toward the generation non-universality, i.e., $\varepsilon_c \neq \varepsilon_u$ and possibly $\varepsilon_s \neq \varepsilon_d$. 

\section{Summaries and Conclusions}

The confirmation of couplings between X17 boson and quarks remains a topic of interest to date. The hypothesis of X17 boson has continuously received attentions by the data of anomalous $^8$Be, $^4$He, and $^{12}$C decays over time. 
The accumulation of data has shifted the most favored coupling strengths from $(\varepsilon_{u}, \varepsilon_{d})= (\pm 3.7 \times 10^{-3}, \mp 7.4\times 10^{-3})$~\cite{Feng:2016jff} to 
$(\varepsilon_u, \varepsilon_d)=(\pm 5\times 10^{-4}, \mp 2.9\times 10^{-3})$ without  isospin effects, and $(\varepsilon_u, \varepsilon_d)=(\pm 9\times 10^{-4}, \mp 2.5\times 10^{-3})$ with both isospin mixing and breaking effects considered~\cite{Denton:2023gat}. 
The persistent decay anomalies outlined above have motivated the testing of X17 hypothesis in the decays of heavy-flavor mesons such as $D^*$ and $B^*$ mesons with the assumption 
$\varepsilon_b=\varepsilon_s=\varepsilon_d$ and $\varepsilon_c=\varepsilon_u$~\cite{Castro:2021gdf}. We have revisited this study by focusing on the comparisons of theoretical predictions (with X17 boson mediated effects included) on decay ratios $R_{D_s^{*+}}$ and  $R_{D^{*0}}$ with experimental data~\cite{CLEO:2011mla,BESIII:2021vyq}. For the case of  $R_{D_s^{*+}}$, we have found that only the updated values of $\varepsilon_{u,d}$ are consistent with the measurement while the earlier extracted parameter values $(\varepsilon_{u}, \varepsilon_{d})= (\pm 3.7 \times 10^{-3}, \mp 7.4\times 10^{-3})$ predict a $R_{D_s^{*+}}$ in a slight  tension with the measurement (see Fig.~\ref{fig:R_D_s}). Our finding corrects an error in Ref.~\cite{Castro:2021gdf} which concluded that the prediction by         
 $(\varepsilon_{u}, \varepsilon_{d})= (\pm 3.7 \times 10^{-3}, \mp 7.4\times 10^{-3})$ is consistent with the measurement. \\
  \indent To accommodate the measured $R_{D^{*0}}$, one requires a much larger $\left|\varepsilon_u\right|$ under the assumption $\varepsilon_c=\varepsilon_u$. On the other hand, raising the magnitude of $\varepsilon_c$ would affect $R_{D_s^{*+}}$ unless the magnitude of $\varepsilon_s$ is increased as well since 
 $R_{D_s^{*+}}$ behaves approximately as $\left| \varepsilon_c+2\varepsilon_s \right|$ with two parameters in opposite signs, as indicated by Fig.~\ref{fig:coup_ucs} (left panel). Since $\varepsilon_s$ already departs from $\varepsilon_d$ in this case,
 we then also  treat $\varepsilon_u$ and $\varepsilon_c$ as independent parameters in our analysis. 
 
 We have predicted the spectrum ${\rm d}R_{D^{*0}}/{\rm d} q^2$ assuming that the excess in $R_{D^{*0}}$ is due to the X17 boson. Our prediction is shown in Fig.~\ref{fig:spectrum} where the $D^{*0}\to D^0 X\to D^0 e^+e^-$ decay peak shows up in the second bin of the spectrum. 
  Since the current statistics of Ref.~\cite{,BESIII:2021vyq} is not yet sufficient to confirm or disconfirm this peak, we have fitted the X17 coupling parameters in two scenarios: one that considers only $R_{D_s^{*+}}$, $R_{\psi(2S)}$, and $R_{\phi}$ data, while the other includes the data of $R_{D^{*0}}$ as well.  
 For the former scenario, the fitting result projected onto  $(\varepsilon_c,\varepsilon_s)$ plane is shown in Fig.~\ref{fig:cs_flavor}.  
 The best-fit values for the coupling parameters are $\left|\varepsilon_{c}\right|=7.6\times 10^{-3}$, $\left|\varepsilon_{s}\right|=2.4\times 10^{-3}$. Although the best-fit value for $\varepsilon_c$ is significantly larger in magnitude than
that of  $\varepsilon_u$ obtained from fitting the ATOMKI data~\cite{Denton:2023gat}, the allowed range for $\varepsilon_c$ is still too large to verify whether these new physics couplings are independent of quark generations or not. 
For the latter scenario, the fitting also yields the range for $\varepsilon_u$ as given in Fig.~\ref{fitting_range}. There are two solutions which correspond to 
the best-fit values of $|\varepsilon_u|$ as $6.1 \times 10^{-2}$ and $6.7 \times 10^{-2}$, respectively. 
The corresponding best-fit couplings, obtained with and without $R_{D^{*0}}$, are summarized in Table~\ref{tab:eps_summary}.

 It is noteworthy that the value of $\left|\varepsilon_u\right|$ shown in Fig.~\ref{fitting_range} remains of the order $10^{-2}$ up to the $3\sigma$ range. Remarkably, this value is much larger than 
 that determined from the anomalous decays of ${ }^{8} \mathrm{Be},{ }^{4} \mathrm{He}$, and ${ }^{12} \mathrm{C}$, i.e., $\left|\varepsilon_{u}\right| =(0.5-0.9) \times 10^{-3}$~\cite{Denton:2023gat}. Such a tension is caused by the large deviation of the measured
 $R_{D^{*0}}$ from the SM prediction as shown in Table~\ref{Tab:Br_Obs}. This is also reflected by the right panel of Fig.~\ref{fig:coup_ucs}, where both green bands significantly deviate from the line $\varepsilon_u=0$. 

We note that the decay channel $D^{*+} \rightarrow D^{+} e^{+} e^{-}$ has not been measured. We have predicted $R_{D^{*+}}$ based on our understanding of X17 related couplings $\varepsilon_d$ and $\varepsilon_c$ as given in Fig.~\ref{fig:RDplus}. 
The central value and the uncertainty of $R_{D^{+*}}$ are both small by assuming the generation universality $\varepsilon_c=\varepsilon_u$.  On the other hand, both the central values and the uncertainties of $R_{D^{+*}}$  become larger if we take the values of $\varepsilon_c$ as those given by the fitting range in Fig.~\ref{fig:cs_flavor}. 
 
 In conclusion, we have studied the effects of the proposed X17 boson to the decays of $D^{*0}$, $D^{*+}_s$, $\psi(2S)$, and $\phi$ mesons, which involve X17 boson couplings to both the first and second generations of quarks. The above-mentioned tension on the value of $\varepsilon_u$ deserves further scrutiny., i.e.,
the huge deviation of the measured $D^{*0} \to D^{0} e^+ e^-$ branching ratio to its SM predicted value awaits confirmations from updated measurements.     

\section*{Acknowledgements}
This work is supported by National Science and Technology Council, Taiwan, under Grant No. NSTC 112-2112-M-A49-017.


\end{document}